\documentclass[reprint,aps]{revtex4-2}

\usepackage{url}
\usepackage{ifthen}
\usepackage[cmex10]{amsmath} 
\usepackage{longtable}
\usepackage{makecell}
\usepackage{algorithm,algorithmic}
                             
\usepackage{amssymb,amsthm}
\usepackage{IEEEtrantools}                           
\usepackage{xifthen,xparse}
\usepackage{hyperref}  
\usepackage{float}
\usepackage{hhline}

\usepackage{tikz,pgfplots,tkz-graph,color}
\pgfplotsset{compat=newest}
\pgfplotsset{plot coordinates/math parser=false}
\usetikzlibrary{decorations.markings,calc,positioning,matrix}

\usepackage{textcomp}

\allowdisplaybreaks

\newtheorem{theorem}{Theorem}
\newtheorem{corollary}[theorem]{Corollary}
\newtheorem{remark}[theorem]{Remark}
\newtheorem{lemma}[theorem]{Lemma}
\newtheorem{example}{Example}

\newenvironment{example*}
  {\addtocounter{example}{-1}\example}
  {\endexample}

\newcommand{\ket}[1]{\left\lvert #1 \right\rangle}
\newcommand{\bra}[1]{\left\langle #1 \right\rvert}
\NewDocumentCommand\ketbra{+m+g}{%
  \IfNoValueTF{#2}
    {\left\lvert #1 \right\rangle \left\langle #1 \right\vert}
  {\left\lvert #1 \right\rangle \left\langle #2 \right\rvert}%
}
\NewDocumentCommand\braket{+m+g}{%
  \IfNoValueTF{#2}
    {\left\langle #1 \vert #1 \right\rangle}
  {\left\langle #1 \vert #2 \right\rangle}%
}

\newcommand{\MCC}{\mathcal{C}}

\newcommand{\vecnot}[1]{\underline{#1}}

\newcommand{\MZ}{\mathbb{Z}}

\newcommand{\syminn}[2]{\langle #1, #2 \rangle_{\text{s}}}

\begin{document}

\title{Unifying the Clifford Hierarchy via Symmetric Matrices over Rings}


\author{Narayanan Rengaswamy}
 \email{narayanan.rengaswamy@duke.edu}
\author{Robert Calderbank}
 \email{robert.calderbank@duke.edu}
\author{Henry D. Pfister}
 \email{henry.pfister@duke.edu}
\affiliation{Department of Electrical and Computer Engineering, Duke University, Durham, North Carolina 27708, USA}

\date{\today}

\begin{abstract}
The Clifford hierarchy of unitary operators is a foundational concept for universal quantum computation.
It was introduced to show that universal quantum computation can be realized via quantum teleportation, given access to certain standard resources.
While the full structure of the hierarchy is still not understood, Cui et al. (Phys. Rev. A \textbf{95}, 012329) recently described the structure of diagonal unitaries in the hierarchy. 
They considered diagonal unitaries whose action on a computational basis qudit state is described by a $2^k$-th root of unity raised to some polynomial function of the state, and they established the level of such unitaries in the hierarchy as a function of $k$ and the degree of the polynomial.
For qubit systems, we consider $k$-th level diagonal unitaries that can be described just by quadratic forms of the state over the ring $\mathbb{Z}_{2^k}$ of integers modulo $2^k$.
The quadratic forms involve symmetric matrices over $\mathbb{Z}_{2^k}$ that can be used to efficiently describe all two-local and certain higher locality diagonal gates in the hierarchy. 
We also provide explicit algebraic descriptions of their action on Pauli matrices, which establishes a natural recursion to diagonal unitaries from lower levels. 
The result involves symplectic matrices over $\mathbb{Z}_{2^k}$ and hence our perspective unifies a subgroup of diagonal gates in the Clifford hierarchy with the binary symplectic framework for gates in the Clifford group.
We augment our description with simple examples for certain standard gates.
In addition to demonstrating structure, these formulas might prove useful in applications such as (i) classical simulation of quantum circuits, especially via the stabilizer rank approach, (ii) synthesis of logical non-Clifford unitaries, specifically alternatives to expensive magic state distillation, and (iii) decomposition of arbitrary unitaries beyond the Clifford+$T$ set of gates, perhaps leading to shorter depth circuits. 
Our results suggest that some non-diagonal gates in the hierarchy might also be understood by generalizing other binary symplectic matrices to integer rings.
\end{abstract}

\maketitle

\section{Introduction}
\label{sec:intro}

Universal quantum computation requires the implementation of arbitrary unitary operators on $m$ qubits.
Gottesman and Chuang showed~\cite{Gottesman-nature99} that universal quantum computation can be achieved via quantum teleportation if one has access to Bell-state preparation, Bell-basis measurements, and arbitrary single-qubit operations on \emph{known} ancilla states.
Their protocol involved construction of the \emph{Clifford hierarchy}.
By definition of the hierarchy, when elements in the $k$-th level act by conjugation on Pauli matrices, they produce a result in the $(k-1)$-th level.
The first level is the Heisenberg-Weyl group of Pauli matrices and the second level is the Clifford group that is fundamental to quantum computation.
It is known that for $k \geq 3$ the unitaries at a level do not form a group~\cite{Zeng-physreva08}. 
The \emph{Gottesman-Knill} theorem~\cite{Gottesman-arxiv98} established that the Clifford group can be efficiently simulated classically and hence does not provide a significant quantum advantage over classical computation (also see~\cite{Aaronson-pra04} for a classical simulator of such circuits).
But the Clifford group combined with \emph{any} unitary outside the group enables arbitrarily good approximation of any other unitary, thus enabling universal quantum computation given the ability to execute a \emph{finite} set of gates~\cite{Boykin-arxiv99}.
The standard choice outside the group is the ``$\pi/8$''- or $T$-gate which belongs to the third level of the Clifford hierarchy.
However, unitaries decomposed with this fixed set of gates could result in circuits with large depth that are especially hard to implement reliably in near-term quantum computers.
It is now established that constant-depth circuits indeed provide a quantum advantage over classical computation~\cite{Bravyi-science18}.
Hence, it is imperative to understand the structure of this hierarchy in order to leverage higher level unitaries and obtain smaller depth circuits.
Moreover, \emph{native} operations in quantum technologies might not belong to the Clifford+$T$ set of gates but to higher levels of the hierarchy, e.g., $X$- and $Z$-rotations of arbitrary angles in trapped-ion systems~\cite{Linke-nas17}. 
Since any circuit must eventually be translated to such native operations by a compiler, this provides us an opportunity to directly consider such operations in circuit decompositions.

There have been several attempts at understanding the structure of the hierarchy~\cite{Zeng-physreva08,Bengtsson-jphysa14,Cui-physreva17}, but the complete structure still remains elusive.
Since the Clifford group is the normalizer of the Pauli group in the unitary group, it permutes maximal commutative subgroups of the Pauli group under conjugation.
Zeng et al.~\cite{Zeng-physreva08} considered a class of unitaries called the \emph{semi-Clifford} operations, which are defined as those unitaries that map \emph{at least one} maximal commutative subgroup of the Pauli group to another maximal commutative subgroup of the Pauli group.
While Gottesman and Chuang~\cite{Gottesman-nature99} used the standard two-ancilla quantum teleportation circuit to demonstrate universal computation, Zhou et al.~\cite{Zhou-pra00} showed that these semi-Clifford operations can be applied via teleportation with \emph{one less} ancilla qubit.
Zeng et al. showed that for $m=1,2$, the unitaries at any level $k$ of the hierarchy are semi-Clifford, and that for $m=3$ all the unitaries in level $k=3$ are semi-Clifford.
For $m > 2$ and $k = 3$, they conjectured that all unitaries are semi-Clifford operations as well, which we believe still remains open.
Furthermore, they also defined \emph{generalized} semi-Clifford operations to be those unitaries that map the span of at least one maximal commutative subgroup of the Pauli group to the span of another maximal commutative subgroup of the Pauli group, where span refers to the group algebra over the complex field.
For $m > 2$ and $k > 3$ they conjectured that all unitaries are generalized semi-Clifford operations but, to the best of our knowledge, this also remains an open problem.

\emph{Stabilizer states} are the unit vectors that belong to the orbit of the computational basis state $\ket{0}^{\otimes m}$ under Clifford operations~\cite{Aaronson-pra04,Dehaene-physreva03}.
Equivalently, they are the common eigenvectors of the commuting Hermitian matrices forming maximal commutative subgroups of the Pauli group.
It is well-known that certain stabilizer states can be grouped and arranged to form mutually unbiased bases (MUBs), which means pairs of vectors within a group are orthogonal and pairs formed from different groups have a small inner product~\cite{Calderbank-seta10,Tirkkonen-isit17}.
The images of stabilizer states under the action of a third level unitary from the Clifford hierarchy are known to produce the states in Alltop's construction of MUBs~\cite{Bengtsson-jphysa14}.
These MUBs are exactly a type of ``magic states'' that provide an alternative path to universal quantum computation~\cite{Bravyi-pra05}.
Bengtsson et al.~\cite{Bengtsson-jphysa14} studied the role of order $3$ Clifford operators, their relation to Alltop MUBs, and a deep connection between Alltop MUBs and symmetric informationally complete (SIC) measurements in quantum mechanics.

The starting point for our contributions is~\cite{Cui-physreva17}, where Cui et al. revealed the structure of the \emph{diagonal} gates in each level of the Clifford hierarchy.
For a single qudit with prime dimension $p$, they constructed a new hierarchy from unitaries of the form $U_{k,a} \triangleq \sum_{j \in \mathbb{Z}_p} \exp\left( \frac{2\pi \imath}{p^k} j^a \right) \ketbra{j}$, where $\mathbb{Z}_p \triangleq \{0,1,\ldots,p-1\}, \imath \triangleq \sqrt{-1}$, and $a$ is an integer such that $1 \leq a \leq p-1$.
They showed that such unitaries determine all diagonal unitaries in the level $(p-1)(k-1)+a$ of the Clifford hierarchy, and they also extended the result to multiple qudits.
In this paper, we provide a simpler description of certain diagonal unitaries (for qubits, i.e., $p=2$) and reveal their structure more explicitly by making a connection to symmetric matrices $R$ over the ring $\MZ_{2^k}$ of integers modulo $2^k$.
We define diagonal unitaries of the form $\tau_R^{(k)} \triangleq \text{diag}\left( \xi^{v R v^T \bmod 2^k} \right) = \sum_{v \in \MZ_2^m} \xi^{v R v^T \bmod 2^k} \ketbra{v}$, where $\xi \triangleq e^{2\pi\imath/2^k}$ and $v$ is a binary (row) vector indexing the rows of the matrix, and prove that all two-local and certain higher locality diagonal unitaries in the $k$-th level can be described in this form (see Theorem~\ref{thm:diagonals} and Remark~\ref{rem:u_ccz}).
We derive precise formulas for their action on Pauli matrices, and show that the result naturally involves a unitary of the form $\tau_{\tilde{R}}^{(k-1)}$, thereby yielding a recursion, where $\tilde{R}$ is a symmetric matrix in $\MZ_{2^{k-1}}$ that is a function of $R$ and the Pauli matrix (see Corollary~\ref{cor:tauR_conjugate}).
Hence the matrix $R$ contains \emph{all} the information about the diagonal unitary $\tau_R^{(k)}$.
Finally, we formally prove that these diagonal unitaries form a subgroup of all diagonal gates in the $k$-th level, and that the map from these unitaries to certain symmetric matrices is an isomorphism.

During this process, we obtain a function $q^{(k-1)}(v; R, a, b)$ (that fully characterizes $\tau_{\tilde{R}}^{(k-1)}$), where $(a,b)$ represents a Pauli matrix (see Section~\ref{sec:prelim}), and we demonstrate some of its properties.
We also provide examples of matrices $R$ for some standard gates, and for the non-Clifford ``$\pi/8$''-gate we clarify the connection between our formula and the well-known action of this gate on the Pauli $X$ matrix.
These symmetric matrices identify symplectic matrices over $\MZ_{2^k}$, and this approach \emph{unifies} these diagonal elements of the Clifford hierarchy with the Clifford group that can be mapped to binary symplectic matrices~\cite{Dehaene-physreva03,Gottesman-arxiv09,Rengaswamy-isit18}. 
We believe this is the first work that provides such a unification, and our results indicate that some non-diagonal unitaries in the Clifford hierarchy might be explored by extending other binary symplectic matrices to rings $\MZ_{2^k}$.

In~\cite{Rengaswamy-isit18}, we exploited the binary symplectic framework for the Clifford group to efficiently assemble all possible physical realizations of a logical Clifford operator for stabilizer codes.
Since, in practice, there might be dynamic hardware constraints such as qubits or qubit links with decreasing fidelity, or non-uniform distributions on the noise, these degrees of freedom might be leveraged to adapt computation to the current environment \emph{without} resorting to codes with large redundancy.
It might be possible to extend this framework to logical (non-Clifford) diagonal unitaries, in a suitable way, using our unification of certain diagonal unitaries with the symplectic representation.
When Paulis are propagated through non-Clifford elements, we lose the Pauli frame, and hence this extension will not be straightforward, but we think research in this direction might produce alternatives to (expensive) magic state distillation~\cite{Bravyi-pra05,Gidney-arxiv18} for realizing non-Clifford logical unitaries.
Moreover, Zeng et al. showed that a semi-Clifford operator $g$ is of the form $g = C_1 D C_2$, where $C_1, C_2$ are Cliffords and $D$ is a diagonal unitary~\cite{Zeng-physreva08}. 
Hence, using calculations similar to those in Section~\ref{sec:discuss} it might be possible to explore the above conjectures by Zeng et al. on semi-Cliffords.
Furthermore, binary symplectic matrices have been used to efficiently decompose Clifford unitaries into circuits composed of standard gates~\cite{Dehaene-physreva03,Can-sen18,Rengaswamy-isit18}.
Using our unification, a better understanding of the interaction between binary and integer symplectic matrices might produce efficient algorithms to decompose unitaries into Cliffords and diagonal gates, thereby also reducing circuit depth.

As another application, classical simulation of quantum circuits is currently an important research topic since it serves at least two purposes: (i) it provides a method to check the integrity of the results produced by near-term quantum computers, and (ii) it refines our understanding of the kind of quantum circuits that indeed provide a computational advantage over classical computation.
Bravyi et al.~\cite{Bravyi-arxiv18} have developed a comprehensive mathematical framework of the notion of \emph{stabilizer rank}, which measures the number of stabilizer states required to express the output state of a given unitary operator, acting on $\ket{0}^{\otimes m}$ without loss of generality.
(Recollect that since Clifford operations can be efficiently simulated classically, each stabilizer state can be easily handled by the \texttt{CHP} simulator of Aaronson and Gottesman~\cite{Aaronson-pra04}, the package on which Bravyi et al. build.)
Using this notion, they have developed a powerful simulator of quantum circuits that can currently handle about $40$-$50$ qubits and over $60$ non-Clifford gates without resorting to high-performance computers.
As they highlight, a key feature of their simulator and a reason for its efficiency is the decomposition of unitaries into Cliffords and arbitrary diagonal gates, such as arbitrary angle $Z$-rotations and controlled-controlled-$Z$ (CCZ) gates, instead of just Cliffords and $T$-gates.
Hence, it is natural to investigate if our symplectic representation of certain diagonal unitaries can be used to extend their simulator.

The paper is organized as follows.
Section~\ref{sec:prelim} introduces notation and background necessary for this work, Section~\ref{sec:diag_cliff} presents the main results, Section~\ref{sec:discuss} discusses potential applications, and finally Section~\ref{sec:conclude} concludes the paper.

\section{Preliminaries}
\label{sec:prelim}

Let $\MZ_{2^k}$ denote the ring of integers modulo $2^k$, for $k \in \mathbb{N}$ (natural numbers), and let $\mathbb{C}$ denote the field of complex numbers.
As a convention we consider vectors over $\MZ_{2^k}$ to be row vectors and vectors over $\mathbb{C}$ to be column vectors.
For $v \in \MZ_2^m$, $e_v = \ket{v}$ denotes the standard basis vector in $\mathbb{C}^N$ with entry $1$ in the position indexed by $v$ and $0$ elsewhere.
Using the binary expansion, we will represent a vector $x \in \MZ^m$ as $x = x_0 + 2x_1 + 4x_2 + \ldots$, where $x_0,x_1,x_2,\ldots \in \MZ_2^m$.
We denote modulo $2$ sums by $\oplus$ and sums in a ring $\MZ_{2^k}$ by $+$.

The single qubit \emph{Pauli} matrices are 
\begin{align}
X \triangleq 
\begin{bmatrix}
0 & 1 \\
1 & 0
\end{bmatrix},\ Z \triangleq 
\begin{bmatrix}
1 & 0 \\
0 & -1
\end{bmatrix},\ Y \triangleq \imath X Z = 
\begin{bmatrix}
0 & -\imath \\
\imath & 0
\end{bmatrix},
\end{align}
and $I_2$, the $2 \times 2$ identity matrix, where $\imath \triangleq \sqrt{-1}$. 
These matrices are unitary and Hermitian.
For $m \in \mathbb{N}$ qubits, let $N \triangleq 2^m$, and define the $N \times N$ matrices
\begin{align}
D(a,b) \triangleq X^{a_1} Z^{b_1} \otimes X^{a_2} Z^{b_2} \otimes \cdots \otimes X^{a_m} Z^{b_m},
\end{align}
where $a = [a_1, a_2, \ldots, a_m], b = [b_1, b_2, \ldots, b_m] \in \mathbb{Z}_2^m$.
Then $D(a,b)^{\dagger} = (-1)^{ab^T} D(a,b)$, 
\begin{align}
E(a,b) \triangleq \imath^{ab^T \bmod 4} D(a,b) = \imath^{ab^T \bmod 4} D(a,0) D(0,b)
\end{align}
is Hermitian and $E(a,b)^2 = I_N$, the $N \times N$ identity matrix.
Note that $D(a,0) = E(a,0)$ are permutation matrices that map $e_v \mapsto e_{v \oplus a}$, and $D(0,b) = E(0,b)$ are diagonal matrices that act like $D(0,b) e_v = (-1)^{vb^T} e_v$. 
Any two such matrices satisfy
\begin{align}
\label{eq:Eab_rule}
E(a,b) E(c,d) & = (-1)^{ad^T + bc^T} E(c,d) E(a,b) \nonumber \\ 
              & = \imath^{bc^T - ad^T} E(a+c, b+d),
\end{align}
where the standard \emph{symplectic inner product} over $\MZ_2^{2m}$ is defined as
\begin{align}
\syminn{[a,b]}{[c,d]} & \triangleq ad^T + bc^T (\bmod\ 2) \nonumber \\
                      & = [a,b]\ \Omega\ [c,d]^T, 
\ \Omega \triangleq
\begin{bmatrix}
0 & I_m \\
I_m & 0
\end{bmatrix}.                      
\end{align}
The \emph{Pauli} or \emph{Heisenberg-Weyl group} $HW_N$ is defined as the group of all matrices $\imath^{\kappa} D(a,b), \kappa \in \mathbb{Z}_4$.

\begin{remark}
\label{rem:general_Eab}
It will be convenient to generalize the above definitions to vectors $x \in \MZ^m$.
Note that this does not distort these definitions since $X^2 = Z^2 = I_2$ implies $D(a,b)$ remains unchanged, while the exponent of $\imath$ for $E(a,b)$ will change to $(a_0+2a_1) (b_0+2b_1)^T = a_0 b_0^T + 2(a_0 b_1^T + a_1 b_0^T)\ (\bmod\ 4)$ which only ever introduces an additional $(-1)$ factor thereby ensuring that $E(a,b)$ is still Hermitian and $E(a,b)^2 = I_N$.
Henceforth all inner (dot) products are performed over $\MZ$, unless mentioned otherwise, and if they occur in the exponent of a $2^k$-th root of unity then the result is automatically reduced modulo $2^k$.
\end{remark}

\begin{table*}
{
\caption{\label{tab:std_symp}\small A generating set of symplectic matrices and their corresponding unitary operators. The number of $1$s in $Q$ and $R$ directly relates to number of gates involved in the circuit realizing the respective unitary operators (see~\cite[Appendix I]{Rengaswamy-isit18}). The $N$ coordinates are indexed by binary vectors $v \in \mathbb{F}_2^m$.
Here $H_{2^t}$ denotes the Walsh-Hadamard matrix of size $2^t$, $U_t = {\rm diag}\left( I_t, 0_{m-t} \right)$ and $L_{m-t} = {\rm diag}\left( 0_t, I_{m-t} \right)$, where $I_t$ is the $t \times t$ identity matrix and $0_t$ is the $t \times t$ matrix with all zero entries. }
\begin{ruledtabular}
\begin{tabular}{ccccc}
~ & ~ & ~ & ~ & ~ \\
Symplectic Matrix $F_g$ & \hspace*{5mm} & Clifford Operator $g$ & \hspace*{5mm} & Circuit Element \\
~ & ~ & ~ & ~ & ~ \\
\hline
~ & ~ & ~ & ~ & ~ \\
$\Omega = \begin{bmatrix} 0 & I_m \\ I_m & 0 \end{bmatrix}$ & \hspace*{5mm} & $H_N = H_2^{\otimes m} = \frac{1}{\sqrt{2^m}} 
\begin{bmatrix}
1 & 1 \\
1 & -1
\end{bmatrix}^{\otimes m}$ & \hspace*{5mm} & Transversal Hadamard \\
~ & ~ & ~ & ~ & ~ \\
$L_Q = \begin{bmatrix} Q & 0 \\ 0 & Q^{-T} \end{bmatrix}$ & \hspace*{5mm} & $\ell_Q: \ket{v} \mapsto \ket{v Q}$ & \hspace*{5mm} & \makecell{Controlled-NOT (CNOT) gates\\ and Permutations} \\
~ & ~ & ~ & ~ & ~ \\
$T_R = \begin{bmatrix} I_m & R \\ 0 & I_m \end{bmatrix}; R = R^T$ & \hspace*{5mm} & $t_R\ =\ {\rm diag}\left( \imath^{v R v^T \bmod 4} \right) = \sum_{v \in \mathbb{F}_2^m} \imath^{v R v^T} \ketbra{v}$ & \hspace*{5mm} &  \makecell{Controlled-$Z$ (CZ) and\\ Phase ($P$) gates}\\
~ & ~ & ~ & ~ & ~ \\
$G_t \Omega^{-1} = \begin{bmatrix} U_t & L_{m-t} \\ L_{m-t} & U_t \end{bmatrix}$ & \hspace*{5mm} & $g_t H_N = I_{2^t} \otimes H_{2^{m-t}}$ & \hspace*{5mm} & Partial Hadamards \\
~ & ~ & ~ & ~ & ~ \\
\end{tabular}
\end{ruledtabular}
%
}
\end{table*}

The first level of the \emph{Clifford hierarchy} is defined to be the Pauli group, i.e., $\MCC^{(1)} \triangleq HW_N$.
The higher levels $k > 1$ of the hierarchy are defined recursively as
\begin{align}
\MCC^{(k)} \triangleq \{ U \in \mathbb{U}_N \colon U D(a,b) U^{\dagger} \in \MCC^{(k-1)} \ \forall \ D(a,b) \in \MCC^{(1)} \},
\end{align}
where $\mathbb{U}_N$ denotes the group of all $N \times N$ unitary matrices~\cite{Gottesman-nature99}.
The second level of the hierarchy $\MCC^{(2)}$ is called the \emph{Clifford group} denoted by $\text{Cliff}_N$.
The Clifford group is the normalizer of the Pauli group in $\mathbb{U}_N$, so elements of $\text{Cliff}_N$ can be mapped to $2m \times 2m$ binary \emph{symplectic} matrices $F$ that preserve the symplectic inner product and hence satisfy $F \Omega F^T = \Omega$ (see~\cite{Rengaswamy-isit18} for a detailed discussion).
Formally, the automorphism induced by a Clifford element $g$ satisfies
\begin{align}
\label{eq:symp_action}
g E(a,b) g^{\dagger} = \pm E\left( [a,b] F_g \right), \ {\rm where} \ \ F_g = 
\begin{bmatrix}
A_g & B_g \\
C_g & D_g
\end{bmatrix}
\end{align}
is symplectic.
The condition $F_g \Omega F_g^T = \Omega$ can be equivalently stated as
$A_g B_g^T = B_g A_g^T, \ C_g D_g^T = D_g C_g^T, \ A_g D_g^T + B_g C_g^T = I_m$.
Let $\text{Sp}(2m,\mathbb{F}_2)$ denote the group of binary symplectic matrices.
The homomorphism $\pi \colon \text{Cliff}_N \rightarrow \text{Sp}(2m,\mathbb{F}_2)$ defined by
$\pi(g) \triangleq F_g $
is surjective with kernel $HW_N$. 
Thus, $HW_N$ is a normal subgroup of $\text{Cliff}_N$ and $\text{Cliff}_N/HW_N \cong \text{Sp}(2m,\mathbb{F}_2)$.
This implies that the size is $|\text{Sp}(2m,\mathbb{F}_2)| = 2^{m^2} \prod_{j=1}^{m} (4^j - 1)$ (also see~\cite{Calderbank-it98*2}).
The symplectic representation is what enables efficient classical simulation of quantum circuits consisting of only Clifford gates~\cite{Gottesman-arxiv98,Aaronson-pra04}.
The elementary symplectic matrices corresponding to standard generators of the Clifford group are shown in Table~\ref{tab:std_symp}.
It is well-known that $\text{Cliff}_N$ combined with any operator from $\mathcal{C}^{(3)}$ enables universal quantum computation.

While each level $k \geq 3$ of the Clifford hierarchy does not form a group, the diagonal unitaries in the $k$-th level of the hierarchy form a group~\cite{Cui-physreva17} that is represented as $\MCC_d^{(k)}$.
We will show that certain elements of $\MCC_d^{(k)}$ can be mapped to symmetric $m \times m$ matrices $R$ over $\MZ_{2^k}$, that in turn determine $2m \times 2m$ matrices $\Gamma = 
\begin{bmatrix}
I_m & R \\
0 & I_m
\end{bmatrix}$ over $\MZ_{2^k}$. 
These also satisfy 
\begin{align}
\label{eq:integer_symp_mat}
\Gamma \Omega \Gamma^T = \Omega\ (\bmod\ 2),
\end{align}
so they are integer symplectic matrices, and hence this generalizes from $\MZ_2$ the third elementary symplectic matrix in Table~\ref{tab:std_symp}.

\section{Diagonal Unitaries in the Clifford Hierarchy}
\label{sec:diag_cliff}

Let $\xi \triangleq \exp\left( \frac{2\pi \imath}{2^k} \right)$ and $R$ be an $m \times m$ \emph{symmetric} matrix over $\MZ_{2^k}$. 
Consider the diagonal unitary matrix
\begin{align}
\tau_R^{(k)} \triangleq \text{diag}\left( \xi^{v R v^T \bmod 2^k} \right) = \sum_{v \in \MZ_2^m} \xi^{v R v^T} \ketbra{v},
\end{align}
where $v \in \MZ_2^m$ indexes the rows of $\tau_R^{(k)}$.
We will derive the action of $\tau_R^{(k)}$ on $E(a,b)$ under conjugation, prove that $\tau_R^{(k)} \in \MCC_d^{(k)}$, and argue that all two-local and certain higher locality diagonal gates can be represented in this form. 
Finally, we will show that the map $\gamma \colon \MCC_{d,\text{sym}}^{(k)} \rightarrow \MZ_{2^k,\text{sym}}^{m \times m}$ defined by $\gamma(\tau_R^{(k)}) \triangleq R$ is an isomorphism, where the subscript ``sym'' denotes symmetric matrices whose diagonal entries are in $\MZ_{2^k}$ and off-diagonal entries are in $\MZ_{2^{k-1}}$, and $\MCC_{d,\text{sym}}^{(k)} \subset \MCC_d^{(k)}$ is the subgroup of all unitaries of the form $\tau_R^{(k)}$. 

Given two vectors $v, w \in \MZ_2^m$, their binary sum can be expressed over $\MZ_{2^k}$ as
\begin{align}
v \oplus w & = v + w - 2 (v \ast w)\ (\bmod\ 2^k),
\end{align}
where $v \ast w$ represents the element-wise product of $v$ and $w$, i.e., $v \ast w = [v_1 w_1, v_2 w_2, \cdots, v_m w_m]$.


\begin{lemma}
\label{lem:binary_quad}
For any $v, w \in \MZ_2^m$, symmetric $R \in \MZ_{2^k}^{m \times m}$, and $k \in \mathbb{N}$, the following holds modulo $2^k$:
\begin{align}
(v \oplus w) R (v \oplus w)^T & \equiv (v + w) R (v + w)^T - 4 \eta(v ; R,w), \\
\text{where}\ \eta(v; R,w) & \triangleq [(v + w) - (v \ast w)] R (v \ast w)^T.
\end{align}
\begin{proof}
We observe that
\begin{align*}
& (v \oplus w) R (v \oplus w)^T \\
  & = [(v + w) - 2(v \ast w)] R [(v + w) - 2(v \ast w)]^T  \\
  & = (v + w) R (v + w)^T - 4 (v + w) R (v \ast w)^T \\
  & \hspace*{5cm} + 4 (v \ast w) R (v \ast w)^T  \\
  & = (v + w) R (v + w)^T - 4 [(v + w) - (v \ast w)] R (v \ast w)^T  \\
  & = (v + w) R (v + w)^T - 4 (v\ \text{OR}\ w) R (v\ \text{AND}\ w)^T  \\
  & = (v + w) R (v + w)^T - 4 \eta(v ; R,w) \ (\bmod\ 2^k).  \tag*{\qedhere}
\end{align*}
\end{proof}
\end{lemma}

For a given binary vector $x$, let $D_x \triangleq \text{diag}(x)$ denote the diagonal matrix with the diagonal set to $x$. 
Then $D_w$ projects onto $w$ so that $D_w v^T = (v \ast w)^T$.
Similarly, $D_{\bar{w}}$ projects onto $\bar{w} = w \oplus \vecnot{1} = \vecnot{1} - w$ so that $v D_{\bar{w}} = v \ast (\vecnot{1} - w) = v - (v \ast w)$, where $\vecnot{1}$ denotes the vector with all entries $1$.
Also, by observing that $v_i^2 = v_i$ for all $i \in \{1,\ldots,m\}$, the inner product $uv^T$ can be expressed as the quadratic form $v D_u v^T$, where $u \in \MZ_2^m$.
Thus, for any $v, w \in \MZ_2^m$, we can write $w R (v \ast w)^T = w R D_w v^T = v D_{wRD_w} v^T$.
It follows that
\begin{align}
\eta(v; R,w) & \triangleq [(v + w) - (v \ast w)] R (v \ast w)^T \nonumber \\
             & = v \, [D_{\bar{w}} R D_w + D_{wRD_w}] \, v^T \nonumber \\
             & = v \, [D_{w} R D_{\bar{w}} + D_{wRD_w}] \, v^T.
\end{align}
%
Next we determine the action of $\tau_R^{(k)}$ on $E(a,b)$ under conjugation (see~\cite[Appendix I-3)]{Rengaswamy-isit18} to compare with the calculation for $t_R \in \text{Cliff}_N$ listed in Table~\ref{tab:std_symp}). 


\begin{lemma}
\label{lem:tauR_conjugate}
Let $k \geq 2, v \in \MZ_2^m, a = a_0 + 2a_1 +4a_2 + \ldots, b = b_0 + 2b_1 + 4b_2 + \ldots,$ and $a_i, b_i \in \MZ_2^m$.
Then,
\begin{align}
\biggr(\tau_R^{(k)} E(a,b) (\tau_R^{(k)})^{\dagger} \biggr) e_v & = \xi^{q^{(k-1)}(v; R,a,b)} E([a_0,b_0] \Gamma_R) e_v \nonumber \\
                                                                & = \xi^{q^{(k-1)}(v; R,a,b)} E(a_0,b_0 + a_0 R) e_v,
\end{align}
where 
$\Gamma_R \triangleq 
\begin{bmatrix}
I_m & R \\
0 & I_m
\end{bmatrix} \in \MZ_{2^k}^{2m \times 2m}$ and 
\begin{align}
q^{(k-1)}(v; R,a,b) & \triangleq (1 - 2^{k-2}) a_0 R a_0^T + 2^{k-1} (a_0 b_1^T + b_0 a_1^T) \nonumber \\
                    & \ \ + (2 + 2^{k-1}) vRa_0^T - 4 \eta(v; R,a_0).
\end{align}
\begin{proof}
We observe $D(a,0) e_v = e_{v \oplus a_0}, D(0,b) e_v = (-1)^{vb_0^T} e_v, \xi^{2^{k-2}} = \imath, \xi^{2^{k-1}} = -1$ and calculate
\begin{widetext}
\begin{align}
\biggr(\tau_R^{(k)} E(a,b) (\tau_R^{(k)})^{\dagger} \biggr) e_v 
  & \overset{\text{(i)}}{=} \imath^{ab^T} \xi^{-vRv^T} \tau_R^{(k)} (-1)^{ab^T} D(0,b) D(a,0) e_v  \\
  & = \imath^{ab^T} \xi^{-vRv^T} (-1)^{a_0 b_0^T} \tau_R^{(k)} (-1)^{(v \oplus a_0)b_0^T} e_{v \oplus a_0}  \\
  & = \imath^{ab^T} \xi^{-vRv^T} (-1)^{a_0 b_0^T} (-1)^{(v + a_0)b_0^T} \xi^{(v \oplus a_0) R (v \oplus a_0)^T} e_{v \oplus a_0}  \\
  & \overset{\text{(ii)}}{=} \xi^{-4 \eta(v; R,a_0)} \imath^{ab^T} (-1)^{a_0 b_0^T} (-1)^{(v + a_0)b_0^T} \xi^{2vRa_0^T + a_0 R a_0^T} e_{v \oplus a_0}  \\
  & \overset{\text{(iii)}}{=} \xi^{a_0 R a_0^T - 4 \eta(v; R,a_0)} \imath^{ab^T} (-1)^{a_0 b_0^T} (-1)^{(v + a_0) (b_0 + a_0 R)^T} (-1)^{a_0 R a_0^T} \xi^{(2 + 2^{k-1}) vRa_0^T} e_{v \oplus a_0}  \\
  & \overset{\text{(iv)}}{=} \xi^{a_0 R a_0^T + (2 + 2^{k-1}) vRa_0^T - 4 \eta(v; R,a_0)} \imath^{ab^T} (-1)^{a_0 (b_0 + a_0 R)^T} D(0, b_0 + a_0 R) D(a_0,0) e_v  \\
  & = \xi^{a_0 R a_0^T + (2 + 2^{k-1}) vRa_0^T - 4 \eta(v; R,a_0)} \imath^{a_0 b_0^T + 2(a_0 b_1^T + b_0 a_1^T)} D(a_0, b_0 + a_0 R) e_v  \\
  & \overset{\text{(v)}}{=} \xi^{(1 - 2^{k-2}) a_0 R a_0^T + 2^{k-1} (a_0 b_1^T + b_0 a_1^T) + (2 + 2^{k-1}) vRa_0^T - 4 \eta(v; R,a_0)} \imath^{a_0 (b_0 + a_0 R)^T} D(a_0, b_0 + a_0R) e_v  \\
  & = \xi^{q^{(k-1)}(v; R,a,b)} E(a_0,b_0 + a_0 R) e_v.
\end{align}
\end{widetext}
In (i), we have applied $(\tau_R^{(k)})^{\dagger}$ to $e_v$ to get the phase $\xi^{-vRv^T}$ and used the fact that $D(a,b) = (-1)^{ab^T} D(0,b) D(0,a)$.
In (ii), we have used Lemma~\ref{lem:binary_quad} to express $(v \oplus a_0) R (v \oplus a_0)^T$ and canceled the factor $\xi^{vRv^T}$ that results with the existing $\xi^{-vRv^T}$.
In (iii), we have rewritten $(v + a_0)b_0^T$ as $(v + a_0)(b_0 + a_0 R)^T - v R a_0^T - a_0 R a_0^T$ and rewritten $(-1)$ as $\xi^{2^{k-1}}$ for the exponent $v R a_0^T$.
In (iv), we have collected all the exponents of $\xi$ and $(-1)$, and then used the fact that $D(0, b_0 + a_0 R) D(a_0,0) e_v = (-1)^{(v + a_0) (b_0 + a_0 R)^T} e_{v \oplus a_0}$.
In (v), we have added and subtracted $a_0 R a_0^T$ in the exponent of $\imath$ and again used the fact that $\xi^{2^{k-2}} = \imath$.
Finally, we have applied the (generalized) definition of $E(a,b)$ (i.e., Remark~\ref{rem:general_Eab}).
\end{proof}
\end{lemma}

\begin{remark}
Consider $k = 2$ so that $\tau_R^{(2)} \in \text{Cliff}_N$ (by Theorem~\ref{thm:diagonals}), and let $a,b \in \MZ_2^m$.
Then we see that $q^{(1)}(v; R,a,b) \equiv 0$ (mod $2^k = 4$), and hence the resulting expression $\tau_R^{(k)} E(a,b) (\tau_R^{(k)})^{\dagger} = E([a,b] \Gamma_R)$ matches exactly with the formula derived for $t_R \in \text{Cliff}_N$ in~\cite[Appendix I-3)]{Rengaswamy-isit18}.
\end{remark}

\begin{example}
Let $m = 1, k = 3$, and consider the ``$\pi/8$''-gate defined by 
$T \triangleq
\begin{bmatrix}
1 & 0 \\
0 & e^{\imath \pi/4}
\end{bmatrix}$.
Since $\xi = e^{\imath \pi/4}$ in this case, it is clear that $R = [\, 1\, ]$.
It is well-known, and direct calculation shows, that $TXT^{\dagger} = \frac{1}{\sqrt{2}} (X + Y)$.
This result can be cast in the form obtained in the above lemma as follows.
For $X = E(1,0)$ we have $a = 1, b = 0$.
So for $v = 0$ we get $q^{(k-1)}(v; R,a,b) = -1$,
\begin{align}
TXT^{\dagger} e_0 & = \tau_R^{(3)} E(1,0) (\tau_R^{(3)})^{\dagger} e_0 \nonumber \\
                        & = \xi^{-1} E(1,0+1) e_0 
                          = e^{-\imath\pi/4} Y e_0.
\end{align}
For $v = 1$ we get $q^{(k-1)}(v; R,a,b) = -1 + 6 - 4 = 1$,
\begin{align}
TXT^{\dagger} e_1 
                        & = \xi^{+1} E(1,0+1) e_1 
                          = e^{\imath\pi/4} Y e_1.
\end{align}
These two actions can be simplified as shown below, where the last steps use $Z e_0 = e_0$ and $Z e_1 = -e_1$. 
\begin{align}
e^{-\imath\pi/4} Y e_0 & = \frac{(1 - \imath)}{\sqrt{2}} Y e_0 = \frac{Y - \imath \times \imath XZ}{\sqrt{2}} e_0 = \frac{Y+X}{\sqrt{2}} e_0, \\
e^{\imath\pi/4} Y e_1 & = \frac{(1 + \imath)}{\sqrt{2}} Y e_1 = \frac{Y + \imath \times \imath XZ}{\sqrt{2}} e_1 = \frac{Y+X}{\sqrt{2}} e_1. 
\end{align}
In this case, the action of $T$ can be unified for both basis vectors $e_0$ and $e_1$ as $\frac{1}{\sqrt{2}} (X+Y)$. \hfill \qedhere
\end{example}


Lemma~\ref{lem:tauR_conjugate} described the result of conjugating a Pauli matrix with a diagonal unitary by its action on the (computational) basis states $e_v$.
It is clear that this action can be expressed without (explicitly writing) these basis states as
\begin{align}
\label{eq:tauR_conjugate}
\tau_R^{(k)} & E(a,b) (\tau_R^{(k)})^{\dagger} \nonumber \\
       & = E([a_0,b_0] \Gamma_R) \, \text{diag}\left( \xi^{q^{(k-1)}(v; R,a,b) \bmod 2^k} \right).
\end{align}
Next we prove a simple corollary that provides a more succinct and recursive description of the above result, using the binary diagonal matrices $D_x$ introduced just before Lemma~\ref{lem:tauR_conjugate}.

\begin{corollary}
\label{cor:tauR_conjugate}
The result of conjugating a Pauli matrix $E(a,b)$ with a diagonal unitary $\tau_R^{(k)}$ can be expressed as
\begin{align}
\tau_R^{(k)} E(a,b) (\tau_R^{(k)})^{\dagger} & = \xi^{\phi(R,a,b,k)} E([a_0,b_0] \Gamma_R) \, \tau_{\tilde{R}(R,a,k)}^{(k-1)},
\end{align}
where the global phase $\phi(R,a,b,k)$ and the new symmetric matrix $\tilde{R}(R,a,k)$ over $\MZ_{2^{k-1}}$ are given by
\begin{align}
\phi(R,a,b,k) & \triangleq (1 - 2^{k-2}) a_0 R a_0^T + 2^{k-1} (a_0 b_1^T + b_0 a_1^T), \\
\tilde{R}(R,a,k) & \triangleq (1 + 2^{k-2}) D_{a_0 R} - (D_{\bar{a}_0} R D_{a_0} + D_{a_0} R D_{\bar{a}_0} \nonumber \\
                 & \hspace*{3.5cm} + 2 D_{a_0 R D_{a_0}}).
\end{align}
Therefore, up to a deterministic global phase, we have
\begin{align}
\label{eq:tau_recursion}
\tau_R^{(k)} E(a,b) (\tau_R^{(k)})^{\dagger} & \equiv E([a_0,b_0] \Gamma_R) \, \tau_{\tilde{R}(R,a,k)}^{(k-1)} \nonumber \\
                                             & = E(a_0,b_0 + a_0 R) \, \tau_{\tilde{R}(R,a,k)}^{(k-1)},
\end{align}
thereby yielding a natural recursion in $k$.
\begin{proof}
Since $vRa_0^T = v \, D_{Ra_0^T} \, v^T = v \, D_{a_0 R} \, v^T$ and $2v \, D_{a_0} R D_{\bar{a}_0} \, v^T = v \, (D_{\bar{a}_0} R D_{a_0} + D_{a_0} R D_{\bar{a}_0}) \, v^T$, we have
\begin{widetext}
\begin{align}
q^{(k-1)}(v; R,a,b) & = (1 - 2^{k-2}) a_0 R a_0^T + 2^{k-1} (a_0 b_1^T + b_0 a_1^T) + (2 + 2^{k-1}) vRa_0^T - 4 \eta(v; R,a_0) \\
  & = (1 - 2^{k-2}) a_0 R a_0^T + 2^{k-1} (a_0 b_1^T + b_0 a_1^T) + (2 + 2^{k-1}) v D_{Ra_0^T} v^T - 4 v [D_{a_0} R D_{\bar{a}_0} + D_{a_0 R D_{a_0}}] v^T \\
  & = (1 - 2^{k-2}) a_0 R a_0^T + 2^{k-1} (a_0 b_1^T + b_0 a_1^T) \nonumber \\
  & \hspace*{3.5cm} + v \, \left[ (2 + 2^{k-1}) D_{a_0 R} - 4 (D_{a_0} R D_{\bar{a}_0} + D_{a_0 R D_{a_0}}) \right] \, v^T \\
  & = (1 - 2^{k-2}) a_0 R a_0^T + 2^{k-1} (a_0 b_1^T + b_0 a_1^T) \nonumber \\
  & \hspace*{3.5cm} + 2 v \, \left[ (1 + 2^{k-2}) D_{a_0 R} - (D_{\bar{a}_0} R D_{a_0} + D_{a_0} R D_{\bar{a}_0} + 2 D_{a_0 R D_{a_0}}) \right] \, v^T \\
  & = \phi(R,a,b,k) + 2 v \, \tilde{R}(R,a,k) \, v^T.
\end{align}
\end{widetext}
Therefore, we can write
\begin{align*}
& \tau_R^{(k)} E(a,b) (\tau_R^{(k)})^{\dagger} \\
  & = E([a_0,b_0] \Gamma_R) \, \text{diag}\left( \xi^{q^{(k-1)}(v; R,a,b) \bmod 2^k} \right) \\
  & = \xi^{\phi(R,a,b,k)} E([a_0,b_0] \Gamma_R) \, \text{diag}\left( (\xi^2)^{v \tilde{R}(R,a,k) v^T \bmod 2^{k-1}} \right) \\
  & = \xi^{\phi(R,a,b,k)} E([a_0,b_0] \Gamma_R) \, \tau_{\tilde{R}(R,a,k)}^{(k-1)}.   \tag*{\qedhere}
\end{align*}
\end{proof}
\end{corollary}

\begin{example*}[contd.]
We have $\phi(R,a,b,k) = -1, \tilde{R}(R,a,k) = [\, 1\, ]$ which implies $TXT^{\dagger} = \xi^{-1} E(1,1) \, \text{diag}(1,\imath) = e^{-\imath\pi/4} Y \, P$. \hfill \qedhere
\end{example*}

\begin{example}
\label{ex:diag_m_1_2}
Consider $m = 1, k = 3$.
The matrices $R$ corresponding to standard single-qubit gates in $\MCC_d^{(3)}$ are:
\begin{IEEEeqnarray*}{rClCrCl}
I_2 & = &
\begin{bmatrix}
1 & 0 \\
0 & 1
\end{bmatrix} \colon R = [ 0 ] & , &\ 
P & = &
\begin{bmatrix}
1 & 0 \\
0 & \imath
\end{bmatrix} \colon R = [ 2 ], \\ 
Z & = &
\begin{bmatrix}
1 & 0 \\
0 & -1
\end{bmatrix} \colon R = [ 4 ] & , &\ 
P^{\dagger} & = &
\begin{bmatrix}
1 & 0 \\
0 & -\imath
\end{bmatrix} \colon R = [ 6 ], \\
T & = &
\begin{bmatrix}
1 & 0 \\
0 & e^{\imath\pi/4} 
\end{bmatrix} \colon R = [ 1 ] & , &\ 
TZ & = &
\begin{bmatrix}
1 & 0 \\
0 & -e^{\imath\pi/4}
\end{bmatrix} \colon R = [ 5 ], \\ 
T^{\dagger} & = &
\begin{bmatrix}
1 & 0 \\
0 & e^{-\imath\pi/4}
\end{bmatrix} \colon R = [ 7 ] & , &\ 
T^{\dagger} Z & = &
\begin{bmatrix}
1 & 0 \\
0 & -e^{-\imath\pi/4} 
\end{bmatrix} \colon R = [ 3 ].
\end{IEEEeqnarray*}
Similarly, for two-qubit gates ($m = 2$) in $\MCC_d^{(3)}$ we have: (C$Z$: Controlled-$Z$, C$P$: Controlled-Phase)
\begin{IEEEeqnarray*}{rCl}
\text{C}Z & = &
\begin{bmatrix}
1 & 0 & 0 & 0 \\
0 & 1 & 0 & 0 \\
0 & 0 & 1 & 0 \\
0 & 0 & 0 & -1
\end{bmatrix} \colon R = 
\begin{bmatrix}
0 & 2 \\
2 & 0
\end{bmatrix}, \\
\text{C}P & = & 
\begin{bmatrix}
1 & 0 & 0 & 0 \\
0 & 1 & 0 & 0 \\
0 & 0 & 1 & 0 \\
0 & 0 & 0 & \imath
\end{bmatrix} \colon R = 
\begin{bmatrix}
0 & 1 \\
1 & 0
\end{bmatrix}, \\ 
I_2 \otimes P & = &
\begin{bmatrix}
1 & 0 & 0 & 0 \\
0 & \imath & 0 & 0 \\
0 & 0 & 1 & 0 \\
0 & 0 & 0 & \imath
\end{bmatrix} \colon R = 
\begin{bmatrix}
0 & 0 \\
0 & 2
\end{bmatrix}, \\
I_2 \otimes Z & = &
\begin{bmatrix}
1 & 0 & 0 & 0 \\
0 & -1 & 0 & 0 \\
0 & 0 & 1 & 0 \\
0 & 0 & 0 & -1
\end{bmatrix} \colon R = 
\begin{bmatrix}
0 & 0 \\
0 & 4
\end{bmatrix}, \\ 
P \otimes I_2 & = &
\begin{bmatrix}
1 & 0 & 0 & 0 \\
0 & 1 & 0 & 0 \\
0 & 0 & \imath & 0 \\
0 & 0 & 0 & \imath
\end{bmatrix} \colon R = 
\begin{bmatrix}
2 & 0 \\
0 & 0
\end{bmatrix}, \\ 
Z \otimes I_2 & = &
\begin{bmatrix}
1 & 0 & 0 & 0 \\
0 & 1 & 0 & 0 \\
0 & 0 & -1 & 0 \\
0 & 0 & 0 & -1
\end{bmatrix} \colon R = 
\begin{bmatrix}
4 & 0 \\
0 & 0
\end{bmatrix}.
\end{IEEEeqnarray*}
\end{example}


Next we prove a simple result that determines the symmetric matrix $R$ for a given diagonal unitary that is a tensor product of diagonal unitaries.

\begin{lemma}
\label{lem:diag_tensor}
Let $\ell, k \in \MZ_{>0}$ such that $\ell < k$, and define $\xi_{\ell} \triangleq \exp(\frac{2\pi \imath}{2^{\ell}}), \xi_k \triangleq \exp(\frac{2\pi \imath}{2^k})$.
Suppose that $\tau_{R_1,m}^{(k)}$ and $\tau_{R_2,n}^{(\ell)}$ are two diagonal unitaries, where $R_1 \in \MZ_{2^k}^{m \times m}$ and $R_2 \in \MZ_{2^{\ell}}^{n \times n}$ are symmetric, and $m,n$ represent the number of qubits on which the unitaries are defined.
Then the symmetric matrix $R \in \MZ_{2^k}^{(m+n) \times (m+n)}$ corresponding to $\tau_{R,m+n}^{(k)} \triangleq \tau_{R_1,m}^{(k)} \otimes \tau_{R_2,n}^{(\ell)}$ is given by $R = 
\begin{bmatrix}
R_1 & 0 \\
0 & 2^{k-\ell} R_2
\end{bmatrix}$.
\begin{proof}
We can simplify the tensor product as follows:
\begin{align*}
& \tau_{R_1,m}^{(k)} \otimes \tau_{R_2,n}^{(\ell)} \\
& = \sum_{v \in \MZ_2^m} \xi_k^{v R_1 v^T \bmod 2^k} \ketbra{v} \otimes \sum_{w \in \MZ_2^n} \xi_{\ell}^{w R_2 w^T \bmod 2^{\ell}} \ketbra{w} \\
& = \sum_{\substack{v \in \MZ_2^m\\w \in \MZ_2^n}} \xi_k^{(v R_1 v^T + 2^{k-\ell} w R_2 w^T) \bmod 2^k} (\ket{v} \otimes \ket{w}) (\bra{v} \otimes \bra{w}) \\
& = \sum_{[v,w] \in \MZ_2^{m+n}} \xi_k^{\small
\begin{bmatrix}
v & w
\end{bmatrix} 
\begin{bmatrix}
R_1 & 0 \\
0 & 2^{k-\ell} R_2
\end{bmatrix}
\begin{bmatrix}
v^T \\ w^T
\end{bmatrix}} \ketbra{v,w} \\
& = \sum_{u \in \MZ_2^{m+n}} \xi_k^{u R u^T} \ketbra{u} 
  = \tau_{R,m+n}^{(k)}.   \tag*{\qedhere}
\end{align*}
\end{proof}
\end{lemma}

The above result can be used to produce the symmetric matrices for the two-qubit tensor product unitaries in Example~\ref{ex:diag_m_1_2} from the symmetric matrices given previously for the single-qubit case.
Now we produce a counterexample for a $3$-local diagonal unitary that cannot be characterized by any symmetric matrix $R$.

\begin{example}
Consider the Controlled-Controlled-$Z$ (CCZ) gate on $m = 3$ qubits represented by the unitary $\text{CC}Z = \text{diag}\, (1,1,1,1,1,1,1,-1)$.
It can be checked that this unitary belongs to level $k = 3$ of the Clifford hierarchy.
Let $R = 
\begin{bmatrix}
a & b & c \\
b & d & e \\
c & e & f
\end{bmatrix}$ be a symmetric matrix with entries in $\MZ_8$.
Equating $\text{CC}Z = \tau_R^{(3)}$, we see that the exponent of $\xi = \exp(\frac{2\pi\imath}{8})$ is $0$ for the first $7$ entries in the diagonal and $-4 \equiv 4$ (mod $8$) for the last entry.
Solving $v R v^T = 0$ for the first $7$ entries, we find that all entries in $R$ have to be $0$.
Thus, there are not enough degrees of freedom in $R$ and we can only produce the identity $I_8$.
\end{example}

Therefore, we have the following result about the diagonal unitaries we characterize in each level of the Clifford hierarchy.

\begin{theorem}
\label{thm:diagonals}
For any symmetric $R \in \MZ_{2^k}^{m \times m}$, the matrix $\tau_R^{(k)} \in \MCC_d^{(k)}$. 
All two-local diagonal unitaries in the Clifford hierarchy can be expressed in the form $\tau_R^{(k)}$ for some $k \in \mathbb{N}$ and symmetric $R \in \MZ_{2^k}^{m \times m}$, up to a global phase.
\begin{proof}
We will prove the first part by induction.
For $k = 1$, $R$ has binary entries and since $\xi = \exp(\frac{2\pi\imath}{2}) = -1$, only the diagonal $d_R$ contributes non-trivially to $vRv^T = \sum_i R_{ii} v_i + 2 \sum_{i < j} R_{ij} v_i v_j$.
So the diagonal entries of $\tau_R^{(1)}$ are $(-1)^{v d_R^T}$ (since $v_i^2 = v_i$), i.e., $\tau_R^{(1)} e_v = (-1)^{v d_R^T} e_v$, and hence $\tau_R^{(1)} = E(0,d_R) \in \MCC_d^{(1)}$. 
Suppose that we have shown $\tau_R^{(k)} \in \MCC_d^{(k)}$ for $k \geq 1$ and any symmetric matrix $R \in \MZ_{2^k}^{m \times m}$.
For level $(k+1)$, we have
\begin{align}
\tau_R^{(k+1)} & E(a,b) (\tau_R^{(k+1)})^{\dagger} \nonumber \\
 & = \xi^{\phi(R,a,b,k+1)} E([a_0,b_0] \Gamma_R) \, \tau_{\tilde{R}(R,a,k+1)}^{(k)}.
\end{align}
Since the global phase can be safely ignored and $\tilde{R}(R,a,k+1) \in \MZ_{2^k}^{m \times m}$ is symmetric, by the induction hypothesis, $\tau_{\tilde{R}(R,a,k+1)}^{(k)} \in \MCC_d^{(k)}$. 
(Note that $\tau_R^{(0)} = I_N$ for all $R$).
Using the fact that the first two levels of the hierarchy are unaffected by multiplication by Paulis, a simple induction shows that if $V \in \MCC^{(k)}$ (not necessarily diagonal) then $E(c,d) V \in \MCC^{(k)}$ as well, for any $c,d \in \MZ_2^m$.
(Note that it is easier to show that $V E(c,d) \in \MCC^{(k)}$ by just using the definition of the hierarchy and the fact that Paulis commute or anti-commute).
Therefore, by the definition of the Clifford hierarchy we have $\tau_R^{(k+1)} \in \MCC_d^{(k+1)}$.
This completes the proof for the first part.

A two-local diagonal unitary $U$ is a tensor product of single- and two-qubit diagonal unitaries.
For $m = 1$, consider a diagonal unitary $W \in \MCC_d^{(k)}$ for any $k \geq 1$.
Then, up to a global phase, there is only one degree of freedom given by the second diagonal entry of $W$ and this must be of the form $\xi^a$ for some $a \in \MZ_{2^k}$~\cite{Cui-physreva17}.
In this case, we can take $R = [\, a\, ]$ so that $W \equiv \tau_R^{(k)}$.
Similarly, for $m = 2$, any diagonal unitary $W$ in the hierarchy has $3$ degrees of freedom with diagonal entries of the form $\xi_k^{\alpha}, \xi_k^{\beta}, \xi_k^{\gamma}$ for some $k \geq 1$, $\xi_k = \exp(\frac{2\pi\imath}{2^k})$, and $\alpha, \beta, \gamma \in \MZ_{2^k}$.
Let $R = 
\begin{bmatrix}
a & b \\
b & c
\end{bmatrix}$ so that the diagonal entries of $\tau_R^{(k)}$ are $\xi_k^c, \xi_k^a, \xi_k^{a+2b+c}$.
Then we can directly set $c = \alpha, a = \beta$ and attempt to solve for $2b = \gamma - a - c$.
If $(\gamma - a - c)$ is even then there exists a $b \in \MZ_{2^k}$, but if $(\gamma - a - c)$ is odd then we can move to level $k + 1$ so that we map $\gamma \mapsto 2 \gamma, a \mapsto 2a, c \mapsto 2c$ (with respect to $\xi_{k+1}$) and then there exists a solution for $b \in \MZ_{2^{k+1}}$.
Hence we satisfy $W \equiv \tau_R^{(\ell)}$ for $\ell = k$ or $k+1$.
Since $U$ is a tensor product of such unitaries, Lemma~\ref{lem:diag_tensor} implies that we can determine the exact symmetric matrix corresponding to $U$.
This completes the proof for the second part.
\end{proof}
\end{theorem}


\begin{example}
Consider the diagonal unitary $U = \text{diag}(1,\imath,\imath,\imath)$.
By the argument in the above proof, we choose $k = 2$ since $\imath = \exp(\frac{2\pi\imath}{2^2})$.
Then using the form of $R$ as in the above proof, we see that $c = a = 1$ given the second and third diagonal entries of $U$.
This implies that we need to find $b$ such that $a + 2b + c = 1 \Rightarrow 2b = -1 \equiv 3$.
Since this does not have a solution in $\MZ_{2^2}$, we move to $k = 3$.
Then we get $c = a = 2$, $2b = 2 - 4 \equiv 6$ and this implies $b = 3$.
Hence, we find that $U = \tau_R^{(3)}$.
\end{example}


\begin{example}
Since we can produce all two-local diagonal unitaries in the hierarchy, we can represent the gate $\text{ZZ}(\theta) \triangleq \exp(-\imath \theta (Z \otimes Z)) = \cos\theta\, I_4 - \imath \sin\theta\, (Z \otimes Z) = \exp(-\imath \theta)\, \text{diag}\, (1, e^{\imath 2\theta}, e^{\imath 2\theta}, 1)$ as $\tau_R^{(k)}$ with $R = 
\begin{bmatrix}
1 & -1 \\
-1 & 1
\end{bmatrix}$, where $\theta = \frac{\pi}{2^k}$ for some $k \geq 1$.
Hence, when combined with Hadamard gates, we can incorporate the M{\o}lmer-S{\o}rensen family of gates $\text{XX}_{ij}(\theta) \triangleq \exp(-\imath \theta\, X_i X_j)$ in our framework, where the subscripts $i$ and $j$ denote the qubits involved in the gate.
Since these gates are the native operations in trapped-ion quantum computers, this observation can potentially lead to applications such as efficient circuit optimization for such systems.
\end{example}

\begin{remark}
\label{rem:u_ccz}
The result in Theorem~\ref{thm:diagonals} only implies that we cannot represent ``all'' $d$-local unitaries for $d > 2$ via a symmetric matrix in our framework.
However, since $\tau_R^{(k)} \in \mathcal{C}_d^{(k)}$ for symmetric $R \in \MZ_{2^k}^{m \times m}$, our framework can generate $2^{mk} 2^{(k-1)m(m-1)/2}$ diagonal gates at the $k$-th level (see Theorem~\ref{thm:symmetric_representation} for the reason behind this count), and this includes a large set of $d$-local unitaries with $d > 2$.
For example, consider the gate $U = \exp(\imath \frac{\pi}{8} (Z \otimes Z \otimes Z)) = \cos\frac{\pi}{8} \, I_8 + \imath \sin\frac{\pi}{8}\, (Z \otimes Z \otimes Z) \in \MCC_d^{(3)}$.
Clearly this gate is $3$-local.
Since $\xi = \exp(\frac{2\pi\imath}{8}) = \exp(\frac{\imath\pi}{4})$, we have $U = \exp(\frac{\imath \pi}{8}) \, \text{diag}\, (\xi^0, \xi^7, \xi^7, \xi^0, \xi^7, \xi^0, \xi^0, \xi^7)$.
Considering $R = 
\begin{bmatrix}
a & b & c \\
b & d & e \\
c & e & f
\end{bmatrix}$ and solving for the entries by setting $v R v^T$ to the above given entries of $U$ (ignoring the global phase), we find that the first seven entries imply $a = d = f = 7, b = c = e = -3 \equiv 5$ (mod $8$).
Therefore, the exponent of the last diagonal entry of $\tau_R^{(3)}$ must be $a + 2b + 2c + d + 2e + f \equiv 3$ whereas the last entry of $U$ is $\xi^7$.
Interestingly, the difference is exactly the factor $\xi^4 = -1$, which means that $\tau_R^{(3)} = U \times \text{CC}Z$ has the above representation $R$ in our framework although it is not a $2$-local unitary.
Note that taking $b = c = e = 1$ does not change the diagonal gate.
\end{remark}

The action of $\tau_R^{(k)}$ on the Pauli matrices directly implies the following result.

\begin{lemma}
\label{lem:isomorphism}
For a fixed $k \in \MZ$ and symmetric $R \in \MZ_{2^k}^{m \times m}$, the map $\varphi \colon E(a,b) \mapsto \tau_R^{(k)} E(a,b) (\tau_R^{(k)})^{\dagger}$ is a group isomorphism.
\end{lemma}
%
%
%

Next we discuss some properties of the objects defined above.


\begin{lemma}
\label{lem:properties}
For $v \in \MZ_2^m$, any $a,b,c,d \in \MZ^m$, and any symmetric $R \in \MZ_{2^k}^{m \times m}$ the following properties hold.
\begin{enumerate}

\item[(a)] The diagonal unitary matrices defined by $\xi$ and $q^{(k-1)}(v; R,a,b)$ satisfy, for any $e,f \in \MZ^m$,
\begin{align}
& \text{diag}\left( \xi^{q^{(k-1)}(v \oplus e_0 ; R,a,b) \bmod 2^k} \right) \nonumber \\
 & = E(e_0,f)\ \text{diag}\left( \xi^{q^{(k-1)}(v; R,a,b) \bmod 2^k} \right) E(e_0,f).
\end{align}

\item[(b)] The function $q^{(k-1)}(v; R, \cdot, \cdot)$ satisfies (modulo $2^k$)
\begin{align}
& q^{(k-1)}(v \oplus c_0; R,a,b) + q^{(k-1)}(v; R,c,d) \nonumber \\
 & = q^{(k-1)}(v; R,a,b) + q^{(k-1)}(v \oplus a_0; R,c,d) \\
  & = q^{(k-1)}(v; R,a+c,b+d) \nonumber \\
  & \qquad + 2^{k-1} (b_0 c_1^T + b_1 c_0^T - a_0 d_1^T - a_1 d_0^T).
\end{align}

\item[(c)] The action of $\tau_R^{(k)}$ satisfies 
\begin{align}
& \tau_R^{(k)} E(c,d) (\tau_R^{(k)})^{\dagger} \times \tau_R^{(k)} E(a,b) (\tau_R^{(k)})^{\dagger} \nonumber \\
 & = E(a_0,e) \left[ \tau_R^{(k)} E(c,d) (\tau_R^{(k)})^{\dagger} \right] E(a_0,e) \nonumber \\
 & \quad \times E(c_0,f) \left[ \tau_R^{(k)} E(a,b) (\tau_R^{(k)})^{\dagger} \right] E(c_0,f),
\end{align}
for any $e,f \in \MZ^m$ such that $\syminn{[a_0,b_0]}{[c_0,d_0]} = \syminn{[a_0,e_0]}{[c_0,f_0]}$, and in particular for $e = b_0 + a_0 R, f = d_0 + c_0 R$.

\end{enumerate}
\begin{proof}
We use identities related to these quantities to complete the proof.
\begin{enumerate}

\item[(a)] Observe that $E(e_0,f) = \imath^{e_0 f^T} E(e_0,0) E(0,f),\ E(0,f) = D(0,f)$ is diagonal and $E(e_0,0) = D(e_0,0)$ is a permutation matrix corresponding to the involution $e_v \mapsto e_{v \oplus e_0}$.

\item[(b)] This can be verified by explicitly enumerating and matching terms on each side of the equality (see~\hyperref[sec:proof_prop1]{Appendix}).
Here we illustrate a more elegant approach.
Using the result of part (a) we calculate
\begin{widetext}
\begin{align}
& \tau_R^{(k)} E(a,b) (\tau_R^{(k)})^{\dagger} \times \tau_R^{(k)} E(c,d) (\tau_R^{(k)})^{\dagger} \nonumber \\
  & = \left[ E([a_0,b_0] \Gamma_R)\ \text{diag}\left( \xi^{q^{(k-1)}(v; R,a,b)} \right) \right] \times \left[ E([c_0,d_0] \Gamma_R)\  \text{diag}\left( \xi^{q^{(k-1)}(v; R,c,d)} \right) \right] \\
  & = E([a_0,b_0] \Gamma_R) E([c_0,d_0] \Gamma_R)\ \text{diag}\left( \xi^{q^{(k-1)}(v \oplus c_0 ; R,a,b)} \right)\ \text{diag}\left( \xi^{q^{(k-1)}(v; R,c,d)} \right) \\
\label{eq:prop2}
  & = (-1)^{\syminn{[a_0,b_0] \Gamma_R}{[c_0,d_0] \Gamma_R}} E([c_0,d_0] \Gamma_R) E([a_0,b_0] \Gamma_R)\ \text{diag}\left( \xi^{q^{(k-1)}(v; R,c,d)} \right)\  \text{diag}\left( \xi^{q^{(k-1)}(v \oplus c_0 ; R,a,b)} \right) \\
\label{eq:prop2b}
  & \overset{\text{(or)}}{=} \imath^{(b_0 + a_0 R) c_0^T - a_0 (d_0 + c_0 R)^T} E([a_0 + c_0, b_0 + d_0] \Gamma_R)\  \text{diag}\left( \xi^{q^{(k-1)}(v \oplus c_0 ; R,a,b)} \right)\ \text{diag}\left( \xi^{q^{(k-1)}(v; R,c,d)} \right). 
\end{align}
\end{widetext}
The first equality uses~\eqref{eq:tauR_conjugate}, the second equality follows from (a), and the last two equalities use the properties given in~\eqref{eq:Eab_rule}.
Note that we have slightly abused notation since the symplectic inner product is defined only for binary vectors.
However, this can be generalized to integer vectors since only their modulo $2$ components play a role in the exponent of $(-1)$.
Once again using the results referenced above, we can also calculate
\begin{widetext}
\begin{align}
& \tau_R^{(k)} E(a,b) (\tau_R^{(k)})^{\dagger} \times \tau_R^{(k)} E(c,d) (\tau_R^{(k)})^{\dagger} \nonumber \\
  & = (-1)^{\syminn{[a_0,b_0]}{[c_0,d_0]}} \tau_R^{(k)} E(c,d) (\tau_R^{(k)})^{\dagger} \times \tau_R^{(k)} E(a,b) (\tau_R^{(k)})^{\dagger} \\
  & = (-1)^{\syminn{[a_0,b_0]}{[c_0,d_0]}} \left[ E([c_0,d_0] \Gamma_R)\ \text{diag}\left( \xi^{q^{(k-1)}(v; R,c,d)} \right) \right] \times \left[ E([a_0,b_0] \Gamma_R)\ \text{diag}\left( \xi^{q^{(k-1)}(v; R,a,b)} \right) \right] \\
  & = (-1)^{\syminn{[a_0,b_0]}{[c_0,d_0]}} E([c_0,d_0] \Gamma_R) E([a_0,b_0] \Gamma_R)\ \text{diag}\left( \xi^{q^{(k-1)}(v \oplus a_0; R,c,d)} \right)\  \text{diag}\left( \xi^{q^{(k-1)}(v ; R,a,b)} \right). 
\end{align}
\end{widetext}
This must be equal to~\eqref{eq:prop2} and, using~\eqref{eq:integer_symp_mat}, we verify
\begin{align}
\syminn{[a_0,b_0] \Gamma_R}{[c_0,d_0] \Gamma_R} & = [a_0,b_0] \Gamma_R\ \Omega\ \Gamma_R^T [c_0, d_0]^T \nonumber \\
%
%
   & = [a_0, b_0] \, \Omega \, [c_0, d_0]^T \nonumber \\
   & = \syminn{[a_0,b_0]}{[c_0,d_0]}
\end{align}
as required (all modulo $2$).
Hence the first equality in the lemma must be true.
Similarly, we have
\begin{align}
& \tau_R^{(k)} E(a,b) (\tau_R^{(k)})^{\dagger} \times \tau_R^{(k)} E(c,d) (\tau_R^{(k)})^{\dagger} \nonumber \\
 & = \tau_R^{(k)} \left[ \imath^{bc^T - ad^T} E(a+c,b+d) \right] (\tau_R^{(k)})^{\dagger} \\
  & = \xi^{2^{k-2}(bc^T - ad^T)} E([a_0 + c_0, b_0 + d_0] \Gamma_R) \nonumber \\
  & \qquad \qquad \times \text{diag}\left( \xi^{q^{(k-1)}(v; R,a+c,b+d)} \right).
\end{align}
Comparing this with~\eqref{eq:prop2b}, and observing that $bc^T - ad^T = b_0 c_0^T - a_0 d_0^T + 2(b_0 c_1^T + b_1 c_0^T - a_0 d_1^T - a_1 d_0^T)\ (\bmod\ 4)$, proves the second equality.

\item[(c)] This follows from the previous properties as shown below.
\begin{widetext}
\begin{align}
& E(a_0,e) \left[ \tau_R^{(k)} E(c,d) (\tau_R^{(k)})^{\dagger} \right] E(a_0,e) \times E(c_0,f) \left[ \tau_R^{(k)} E(a,b) (\tau_R^{(k)})^{\dagger} \right] E(c_0,f) \\
 & = E(a_0,e) E(c_0,d_0 + c_0 R)\ \text{diag}\left( \xi^{q^{(k-1)}(v; R,c,d)} \right) E(a_0,e) \nonumber \\
 & \hspace*{3.5cm}  \times E(c_0,f) E(a_0,b_0 + a_0 R)\ \text{diag}\left( \xi^{q^{(k-1)}(v; R,a,b)} \right) E(c_0,f) \\
 & = (-1)^{a_0 (d_0 + c_0 R)^T + e_0 c_0^T} E(c_0, d_0 + c_0 R)\ \text{diag}\left( \xi^{q^{(k-1)}(v \oplus a_0; R,c,d)} \right) \nonumber \\
 & \hspace*{3.5cm} \times (-1)^{c_0 (b_0 + a_0 R)^T + f_0 a_0^T} E(a_0, b_0 + a_0 R)\ \text{diag}\left( \xi^{q^{(k-1)}(v \oplus c_0; R,a,b)} \right)  \\
 & = (-1)^{\syminn{[a_0,b_0]}{[c_0,d_0]} + \syminn{[a_0,e_0]}{[c_0,f_0]}} E(c_0, d_0 + c_0 R) E(a_0, b_0 + a_0 R)\, \text{diag}\left( \xi^{q^{(k-1)}(v; R,c,d)} \right) \text{diag}\left( \xi^{q^{(k-1)}(v \oplus c_0; R,a,b)} \right) \\
 & = E(c_0, d_0 + c_0 R) E(a_0, b_0 + a_0 R) \ \text{diag}\left( \xi^{q^{(k-1)}(v \oplus a_0; R,c,d)} \right) \text{diag}\left( \xi^{q^{(k-1)}(v; R,a,b)} \right) \\
 & = E(c_0, d_0 + c_0 R)\ \text{diag}\left( \xi^{q^{(k-1)}(v; R,c,d)} \right) \times E(a_0, b_0 + a_0 R)\ \text{diag}\left( \xi^{q^{(k-1)}(v; R,a,b)} \right) \\
 & = \tau_R^{(k)} E(c,d) (\tau_R^{(k)})^{\dagger} \times \tau_R^{(k)} E(a,b) (\tau_R^{(k)})^{\dagger}. 
\end{align}
\end{widetext}
Again, the first equality uses~\eqref{eq:tauR_conjugate}. 
The second equality uses the properties in~\eqref{eq:Eab_rule} to swap the order of Paulis, then uses the result of (a) to pass $E(a_0,e)$ and $E(c_0,f)$ through the diagonals, and then observes the property that $E(a_0,e)^2 = E(c_0,f)^2 = I_N$.
The third equality collects exponents by noting that $a_0 R c_0^T = c_0 R a_0^T$ (since $R$ is symmetric), and then uses the result of (a) to pass $E(a_0, b_0 + a_0 R)$ through the diagonal on its left.
The fourth equality utilizes the condition assumed in the hypothesis as well as the result of (b).
The fifth equality once again uses (a) to pass back $E(a_0, b_0 + a_0 R)$, and finally the last step follows from~\eqref{eq:tauR_conjugate}.

\end{enumerate}
This completes the proof. \hfill \qedhere
\end{proof}
\end{lemma}




\begin{theorem}
\label{thm:symmetric_representation}
Fix $k \geq 1$.
Define $\MCC_{d,\text{sym}}^{(k)}$ to be the set of diagonal unitaries $\tau_R^{(k)}$ for all matrices $R \in \MZ_{2^k, \text{sym}}^{m \times m}$, where the subscript ``sym'' represents symmetric matrices whose diagonal entries are in $\MZ_{2^k}$ and off-diagonal entries are in $\MZ_{2^{k-1}}$.
Then $\MCC_{d,\text{sym}}^{(k)}$ is a subgroup of $\MCC_d^{(k)}$.
Furthermore, the map $\gamma \colon \MCC_{d,\text{sym}}^{(k)} \rightarrow \MZ_{2^k,\text{sym}}^{m \times m}$ defined by $\gamma(\tau_{R}^{(k)}) \triangleq R$ is an isomorphism. 
\begin{proof}
From Theorem~\ref{thm:diagonals} we know that $\tau_R^{(k)} \in \MCC_d^{(k)}$. 
Then 
\begin{align}
\gamma\left( \tau_{R_1}^{(k)} \times \tau_{R_2}^{(k)} \right) & = \gamma\left(\text{diag}\left( \xi^{vR_1v^T} \right) \times \text{diag}\left( \xi^{vR_2v^T} \right) \right) \\
  & = \gamma\left( \tau_{R_1 + R_2}^{(k)} \right) \\
  & = R_1 + R_2 \\
  & = \gamma\left( \tau_{R_1}^{(k)} \right) + \gamma\left( \tau_{R_2}^{(k)} \right).
\end{align}
As discussed in the proof of Theorem~\ref{thm:diagonals}, since $vRv^T = \sum_i R_{ii} v_i + 2 \sum_{i < j} R_{ij} v_i v_j$, when $2^{k-1}$ is added to any off-diagonal entry $R_{ij}$, the factor of $2$ produces $2^k R_{ij} v_i v_j$ which vanishes modulo $2^k$ (see Remark~\ref{rem:u_ccz} for an example).
Therefore, only when the off-diagonal entries are restricted to values in the ring $\MZ_{2^{k-1}}$, the vectors $[v R_1 v^T]_{v \in \MZ_2^m}$ and $[v R_2 v^T]_{v \in \MZ_2^m}$ are distinct for distinct $R_1, R_2$ and $k \geq 1$. 
Here, the sum $R_1 + R_2$ is taken over $\MZ_{2^k}$ for the diagonal entries and over $\MZ_{2^{k-1}}$ for the off-diagonal entries.
Hence, the closure implies that $\MCC_{d,\text{sym}}^{(k)}$ is clearly a subgroup of $\MCC_d^{(k)}$. 
Moreover, by definition $\MCC_{d,\text{sym}}^{(k)}$ does not include global phases, so the map $\gamma$ is an isomorphism.
\end{proof} 
\end{theorem}

\section{Discussion}
\label{sec:discuss}

In this section, we describe how we might apply our new characterization to classical simulation of quantum circuits, synthesis of logical diagonal unitaries, and decomposition of unitaries into Cliffords and diagonal gates.

The classical simulation problem can be succinctly described as follows.
Given a unitary operator $U$ acting on $\ket{0}^{\otimes m}$ to produce the state $\ket{\psi} = U \ket{0}^{\otimes m}$, efficiently sample from the distribution $P_{\psi}(x) = |\braket{x}{\psi}|^2$, where $x \in \mathbb{Z}_2^m$.
We know that the stabilizer for the initial state $\ket{0}^{\otimes m}$ is $Z_N \triangleq \{ E(0,b) \colon b \in \mathbb{Z}_2^m \}$.
Note that this is a maximal commutative subgroup of the Pauli group as it has $m$ generators.
If $U \in \text{Cliff}_N$, we can track the stabilizer of the state $\ket{\psi}$ as $U Z_N U^{\dagger}$, which can be done efficiently using the symplectic representation of $U$ and the identity~\eqref{eq:symp_action}.
More generally, any unitary $U$ can be decomposed as
\begin{align}
U = C_n D_n C_{n-1} D_{n-1} \cdots C_1 D_1 C_0,
\end{align}
where $C_i \in \text{Cliff}_N$ and $D_i \in \mathcal{C}_d^{(k_i)}$ for $k_i \in \{3,4,\ldots\}$~\cite{Bravyi-arxiv18}.
For simplicity, assume $k_i = k$ for all $i$. 
First, let $n = 1$ and let the stabilizer before $C_0$ be $S = \langle E(a_j,b_j); j=1,\ldots,m \rangle$ to keep the initial state generic.
(Each $E(a_j,b_j)$ can also have an overall $(-1)$ factor, but we ignore this since it does not provide any new insight.)
Let $F_0$ be the symplectic matrix corresponding to $C_0$.
Then the new stabilizer can be expressed as
\begin{align}
S_0 & = \langle C_0 E(a_j,b_j) C_0^{\dagger}; j=1,\ldots,m \rangle \\
    & = \langle \pm E([a_j,b_j]F_0); j=1,\ldots,m \rangle.
\end{align}
The \texttt{CHP} simulator of Aaronson and Gottesman~\cite{Aaronson-pra04} indeed keeps track of the stabilizer in this manner and the stabilizer rank approach of Bravyi et al. builds on this~\cite{Bravyi-arxiv18}.
Define $[a_{0,j}, b_{0,j}] \triangleq [a_j,b_j]F_0$.
Suppose $D_1 = \tau_{R_1}^{(k)}$ for some symmetric $R_1$ and let $\Gamma_1 = 
\begin{bmatrix}
I_m & R_1 \\
0 & I_m
\end{bmatrix}$. 
Then, using Corollary~\ref{cor:tauR_conjugate}, we can track the new stabilizer after $D_1$ as
\begin{align}
S_1' &= \langle \pm \tau_{R_1}^{(k)} E(a_{0,j},b_{0,j}) (\tau_{R_1}^{(k)})^{\dagger}; j=1,\ldots,m \rangle \\
  & = \langle \pm \xi^{\phi(R_1,a_{0,j},b_{0,j},k)} E([a_j,b_j]F_0 \Gamma_1) \nonumber \\
  & \qquad \qquad \quad \ \times \tau_{\tilde{R}_1(R_1,a_{0,j},k)} ; j=1,\ldots,m \rangle.
\end{align}
At this point, note that each stabilizer generator is completely determined by $a_j,b_j,F_0$ and $\Gamma_1$ (or equivalently $R_1$), whose sizes grow only as $O(m^2)$.
Next, let $F_1$ be the binary symplectic matrix corresponding to $C_1$.
Then the new stabilizer is
\begin{align}
S_1 & = \langle \pm \xi^{\phi(R_1,a_{0,j},b_{0,j},k)} C_1 E([a_j,b_j]F_0 \Gamma_1) C_1^{\dagger} \nonumber \\
    & \hspace*{1.6cm} \times C_1 \tau_{\tilde{R}_1(R_1,a_{0,j},k)} C_1^{\dagger} ; j=1,\ldots,m \rangle \\
  & = \langle \pm \xi^{\phi(R_1,a_{0,j},b_{0,j},k)} E([a_j,b_j]F_0 \Gamma_1 F_1) \nonumber \\
  & \qquad \times \left( C_1 \tau_{\tilde{R}_1(R_1,a_{0,j},k)} C_1^{\dagger} \right) ; j=1,\ldots,m \rangle.
\end{align}
We could expand the second term in each generator as follows.
For simplicity, just consider some $g \in \text{Cliff}_N$ and a $\tau_R^{(k)} \in \mathcal{C}_d^{(k)}$.
\begin{align}
g \tau_R^{(k)} g^{\dagger} & = g \left( \sum_{v \in \mathbb{Z}_2^m} \xi^{vRv^T \bmod 2^k} \ketbra{v} \right) g^{\dagger} \\
                           & = \sum_{v \in \mathbb{Z}_2^m} \xi^{vRv^T \bmod 2^k} g \ketbra{v} g^{\dagger}.
\end{align}
So now the stabilizer involves operators that are diagonal in an eigenbasis of stabilizer states $\{ g\ket{v} \}$.
If we proceed as before to apply another diagonal gate $D_2$ then the interactions become more complicated as we might expect, since arbitrary stabilizers are indeed hard to track and this is one way to see the gap between quantum and classical computation.
However, we see that our perspective enables to continue this recursion and shows that every stabilizer generator is \emph{structured}: it always involves a Hermitian Pauli matrix, that can be \emph{efficiently} tracked using the symplectic matrices $F_i$ and $\Gamma_i$, and additional terms that become more complex with the depth of the decomposition of $U$.

Although we did this calculation in the context of classical simulation, it captures the calculations in the other two applications as well.
For logical Clifford operations, once we generate logical Paulis using Gottesman's~\cite{Gottesman-phd97} or Wilde's~\cite{Wilde-physreva09} algorithm, we need to perform the above type of calculations to impose linear constraints on the target symplectic matrix that represents the physical realization of the logical operator (see~\cite{Rengaswamy-isit18} for details).
Although the same approach can be attempted for logical diagonal unitaries, the fact that we need to fix the code by normalizing the stabilizer introduces complications.
In other words, when the (Pauli) stabilizer of the code is conjugated by a non-Clifford operator, the stabilizer generators are no more purely Paulis and hence the code space might be disturbed.
This is the challenge overcome by magic state distillation~\cite{Bravyi-pra05}, but since that procedure is usually expensive, we think it will be interesting to explore if our unification via symplectic matrices produces alternative strategies for non-Clifford (diagonal) logical operations.
Similarly, Clifford unitaries are decomposed by suitably multiplying elementary symplectic matrices from Table~\ref{tab:std_symp} (see~\cite{Dehaene-physreva03},\cite[Appendix I]{Rengaswamy-isit18}).
In order to produce decompositions of the form shown above for a general unitary $U$, we need to understand the interaction between binary symplectic matrices $F_i$ and integer symplectic matrices $\Gamma_i$.
Such an understanding might enable us to develop decomposition algorithms that take advantage of \emph{native} operations in quantum technologies such as arbitrary angle $X$- and $Z$-rotations, and M{\o}lmer-S{\o}rensen gates, in trapped-ion architectures~\cite{Linke-nas17}.
For these purposes, it will be interesting to see if the properties described in Lemma~\ref{lem:properties} can be effectively put to use.


\section{Conclusion}
\label{sec:conclude}

In this work we provided a simpler description of certain diagonal gates in the Clifford hierarchy, and derived explicit formulas for their action on Pauli matrices.
We established an isomorphism between these unitaries and certain symmetric matrices over rings $\MZ_{2^k}$ that carries all information about the unitaries. 
These symmetric matrices further determine symplectic matrices over $\MZ_{2^k}$, thereby providing a natural generalization to the mapping of Clifford group elements to binary symplectic matrices.
It remains to be explored if our explicit characterization can be used to improve classical simulation of certain classes of quantum circuits, synthesis of logical diagonal unitaries, and decomposition of generic unitaries into Cliffords and diagonal gates.
Another interesting open problem is whether some non-diagonal elements of the Clifford hierarchy can be understood by generalizing other standard binary symplectic matrices to rings $\MZ_{2^k}$.

\begin{acknowledgments}
We would like to thank Theodore Yoder for pointing out that our framework does not include all $d$-local diagonal gates, for $d > 2$, as we had originally thought.
He provided the counterexample in Remark~\ref{rem:u_ccz}.

The work of H. D. Pfister and N. Rengaswamy was supported in part by the National Science Foundation (NSF) under Grant No. 1718494. Any opinions, findings, conclusions, and recommendations expressed in this material are those of the authors and do not necessarily reflect the views of these sponsors.
\end{acknowledgments}

\appendix*

\section{Alternate Proof of Lemma~\ref{lem:properties}(b)}
\label{sec:proof_prop1}

We ignore the common terms $q^{(k-1)}(v; R,a,b) + q^{(k-1)}(v; R,c,d)$ on both sides of the equality and consider only the remaining terms.
Note that the calculation is modulo $2^k$.
Let $\tilde{c}_0 = c_0 - 2 (v \ast c_0)$.
For the left hand side we have, by first ignoring $q^{(k-1)}(v; R,c,d)$ and subsequently $q^{(k-1)}(v; R,a,b)$,
\newpage
\begin{widetext}
\begin{align}
& q^{(k-1)}(v \oplus c_0; R,a,b) \nonumber \\
& = (1 - 2^{k-2}) a_0 R a_0^T + 2^{k-1} (a_0 b_1^T + b_0 a_1^T) + (2 + 2^{k-1}) (v \oplus c_0) Ra_0^T \nonumber \\
& \hspace*{9cm} - 4 [((v \oplus c_0) + a_0) - ((v \oplus c_0) \ast a_0)] R ((v \oplus c_0) \ast a_0)^T \\
& = (1 - 2^{k-2}) a_0 R a_0^T + 2^{k-1} (a_0 b_1^T + b_0 a_1^T) + (2 + 2^{k-1}) (v + \tilde{c}_0) Ra_0^T \nonumber \\
& \hspace*{9cm} - 4 [((v + \tilde{c}_0) + a_0) - ((v + \tilde{c}_0) \ast a_0)] R ((v + \tilde{c}_0) \ast a_0)^T \\
& = q^{(k-1)}(v; R,a,b) + (2 + 2^{k-1}) \tilde{c}_0 R a_0^T - 4 \biggr[ (v + a_0 - v \ast a_0) R (\tilde{c}_0 \ast a_0)^T + (\tilde{c}_0 - \tilde{c}_0 \ast a_0) R (v \ast a_0)^T \biggr] \nonumber \\
 & \hspace*{11.5cm} + (\tilde{c}_0 - \tilde{c}_0 \ast a_0) R (\tilde{c}_0 \ast a_0)^T \biggr] \\
& \equiv (2 + 2^{k-1}) c_0 R a_0^T - 4 (v \ast c_0) R a_0^T - 4 (v + a_0 - v \ast a_0) R (c_0 \ast a_0)^T + 8 (v + a_0 - v \ast a_0) R (v \ast c_0 \ast a_0)^T \nonumber \\
& \quad - 4 (c_0 - 2(v \ast c_0)) R (v \ast a_0)^T + 4 ((c_0 \ast a_0) - 2 v \ast c_0 \ast a_0) R (v \ast a_0)^T - 4 (c_0 - 2 v \ast c_0) R ((c_0 - 2 v \ast c_0) \ast a_0)^T \nonumber \\
& \quad + 4 (c_0 \ast a_0 - 2 v \ast c_0 \ast a_0) R (c_0 \ast a_0 - 2 v \ast c_0 \ast a_0)^T \\
& = [(2 + 2^{k-1}) c_0 R a_0^T]_1 - [4 (v \ast c_0) R a_0^T]_2 - [4 v R (c_0 \ast a_0)^T]_3 - [4 a_0 R (c_0 \ast a_0)^T]_4 + [4 (v \ast a_0) R (c_0 \ast a_0)^T]_5 \nonumber \\
& \quad + [8 v R (v \ast c_0 \ast a_0)^T]_6 + [8 a_0 R (v \ast c_0 \ast a_0)^T]_7 - [8 (v \ast a_0) R (v \ast c_0 \ast a_0)^T]_8 - [4 c_0  R (v \ast a_0)^T]_2 \nonumber \\
& \quad + [8 (v \ast c_0)  R (v \ast a_0)^T]_9 + [4 (c_0 \ast a_0) R (v \ast a_0)^T]_5 - [8 (v \ast c_0 \ast a_0) R (v \ast a_0)^T]_8 - [4 c_0 R (c_0 \ast a_0)^T]_4 \nonumber \\
& \quad + [8 c_0 R (v \ast c_0 \ast a_0)^T]_7 + [8 (v \ast c_0) R (c_0 \ast a_0)^T]_5 - [16 (v \ast c_0) R (v \ast c_0 \ast a_0)^T]_8 + [4 (c_0 \ast a_0) R (c_0 \ast a_0)^T]_{10} \nonumber \\
& \quad - [16 (c_0 \ast a_0) R (v \ast c_0 \ast a_0)^T]_{11} + [16 (v \ast c_0 \ast a_0) R (v \ast c_0 \ast a_0)^T]_{12}.
\end{align}
\end{widetext}
Observe that using the same strategy as above, the terms for the right hand side (of the first equality in Lemma~\ref{lem:properties}(b)) will simply be the above expression with $a_0$ and $c_0$ swapped.
The numbers in the subscript are given to facilitate matching the terms obtained by swapping $a_0$ and $c_0$.
A quick inspection shows that every term is either symmetric about $a_0$ and $c_0$ or has a pair under the swap, and hence the overall expression remains the same.
Therefore the two sides are equal and this completes the proof of the first equality in Lemma~\ref{lem:properties}(b). \hfill \qed


\begin{thebibliography}{22}%
\makeatletter
\providecommand \@ifxundefined [1]{%
 \@ifx{#1\undefined}
}%
\providecommand \@ifnum [1]{%
 \ifnum #1\expandafter \@firstoftwo
 \else \expandafter \@secondoftwo
 \fi
}%
\providecommand \@ifx [1]{%
 \ifx #1\expandafter \@firstoftwo
 \else \expandafter \@secondoftwo
 \fi
}%
\providecommand \natexlab [1]{#1}%
\providecommand \enquote  [1]{``#1''}%
\providecommand \bibnamefont  [1]{#1}%
\providecommand \bibfnamefont [1]{#1}%
\providecommand \citenamefont [1]{#1}%
\providecommand \href@noop [0]{\@secondoftwo}%
\providecommand \href [0]{\begingroup \@sanitize@url \@href}%
\providecommand \@href[1]{\@@startlink{#1}\@@href}%
\providecommand \@@href[1]{\endgroup#1\@@endlink}%
\providecommand \@sanitize@url [0]{\catcode `\\12\catcode `\$12\catcode
  `\&12\catcode `\#12\catcode `\^12\catcode `\_12\catcode `\%12\relax}%
\providecommand \@@startlink[1]{}%
\providecommand \@@endlink[0]{}%
\providecommand \url  [0]{\begingroup\@sanitize@url \@url }%
\providecommand \@url [1]{\endgroup\@href {#1}{\urlprefix }}%
\providecommand \urlprefix  [0]{URL }%
\providecommand \Eprint [0]{\href }%
\providecommand \doibase [0]{https://doi.org/}%
\providecommand \selectlanguage [0]{\@gobble}%
\providecommand \bibinfo  [0]{\@secondoftwo}%
\providecommand \bibfield  [0]{\@secondoftwo}%
\providecommand \translation [1]{[#1]}%
\providecommand \BibitemOpen [0]{}%
\providecommand \bibitemStop [0]{}%
\providecommand \bibitemNoStop [0]{.\EOS\space}%
\providecommand \EOS [0]{\spacefactor3000\relax}%
\providecommand \BibitemShut  [1]{\csname bibitem#1\endcsname}%
\let\auto@bib@innerbib\@empty
\bibitem [{\citenamefont {Gottesman}\ and\ \citenamefont
  {Chuang}(1999)}]{Gottesman-nature99}%
  \BibitemOpen
  \bibfield  {author} {\bibinfo {author} {\bibfnamefont {D.}~\bibnamefont
  {Gottesman}}\ and\ \bibinfo {author} {\bibfnamefont {I.~L.}\ \bibnamefont
  {Chuang}},\ }\bibfield  {title} {\bibinfo {title} {{Demonstrating the
  viability of universal quantum computation using teleportation and
  single-qubit operations}},\ }\href {http://www.nature.com/articles/46503}
  {\bibfield  {journal} {\bibinfo  {journal} {Nature}\ }\textbf {\bibinfo
  {volume} {402}},\ \bibinfo {pages} {390} (\bibinfo {year}
  {1999})}\BibitemShut {NoStop}%
\bibitem [{\citenamefont {Zeng}\ \emph {et~al.}(2008)\citenamefont {Zeng},
  \citenamefont {Chen},\ and\ \citenamefont {Chuang}}]{Zeng-physreva08}%
  \BibitemOpen
  \bibfield  {author} {\bibinfo {author} {\bibfnamefont {B.}~\bibnamefont
  {Zeng}}, \bibinfo {author} {\bibfnamefont {X.}~\bibnamefont {Chen}},\ and\
  \bibinfo {author} {\bibfnamefont {I.~L.}\ \bibnamefont {Chuang}},\ }\bibfield
   {title} {\bibinfo {title} {{Semi-Clifford operations, structure of
  $\mathcal{C}_k$ hierarchy, and gate complexity for fault-tolerant quantum
  computation}},\ }\href
  {https://journals.aps.org/pra/abstract/10.1103/PhysRevA.77.042313} {\bibfield
   {journal} {\bibinfo  {journal} {Phys. Rev. A}\ }\textbf {\bibinfo {volume}
  {77}},\ \bibinfo {pages} {042313} (\bibinfo {year} {2008})},\ \bibinfo {note}
  {[Online]. Available: http://arxiv.org/abs/0712.2084}\BibitemShut {NoStop}%
\bibitem [{\citenamefont {Gottesman}(1998)}]{Gottesman-arxiv98}%
  \BibitemOpen
  \bibfield  {author} {\bibinfo {author} {\bibfnamefont {D.}~\bibnamefont
  {Gottesman}},\ }\bibfield  {title} {\bibinfo {title} {{The Heisenberg
  Representation of Quantum Computers}},\ }\href@noop {} {\bibfield  {journal}
  {\bibinfo  {journal} {arXiv preprint arXiv:quant-ph/9807006}\ } (\bibinfo
  {year} {1998})},\ \bibinfo {note} {[Online]. Available:
  https://arxiv.org/pdf/quant-ph/9807006.pdf}\BibitemShut {NoStop}%
\bibitem [{\citenamefont {Aaronson}\ and\ \citenamefont
  {Gottesman}(2004)}]{Aaronson-pra04}%
  \BibitemOpen
  \bibfield  {author} {\bibinfo {author} {\bibfnamefont {S.}~\bibnamefont
  {Aaronson}}\ and\ \bibinfo {author} {\bibfnamefont {D.}~\bibnamefont
  {Gottesman}},\ }\bibfield  {title} {\bibinfo {title} {{Improved simulation of
  stabilizer circuits}},\ }\href
  {https://journals.aps.org/pra/pdf/10.1103/PhysRevA.70.052328} {\bibfield
  {journal} {\bibinfo  {journal} {Phys. Rev. A}\ }\textbf {\bibinfo {volume}
  {70}},\ \bibinfo {pages} {052328} (\bibinfo {year} {2004})}\BibitemShut
  {NoStop}%
\bibitem [{\citenamefont {Boykin}\ \emph {et~al.}(1999)\citenamefont {Boykin},
  \citenamefont {Mor}, \citenamefont {Pulver}, \citenamefont {Roychowdhury},\
  and\ \citenamefont {Vatan}}]{Boykin-arxiv99}%
  \BibitemOpen
  \bibfield  {author} {\bibinfo {author} {\bibfnamefont {P.~O.}\ \bibnamefont
  {Boykin}}, \bibinfo {author} {\bibfnamefont {T.}~\bibnamefont {Mor}},
  \bibinfo {author} {\bibfnamefont {M.}~\bibnamefont {Pulver}}, \bibinfo
  {author} {\bibfnamefont {V.}~\bibnamefont {Roychowdhury}},\ and\ \bibinfo
  {author} {\bibfnamefont {F.}~\bibnamefont {Vatan}},\ }\bibfield  {title}
  {\bibinfo {title} {{On Universal and Fault-Tolerant Quantum Computing}},\
  }\href@noop {} {\bibfield  {journal} {\bibinfo  {journal} {arXiv preprint
  arXiv:quant-ph/9906054}\ } (\bibinfo {year} {1999})},\ \bibinfo {note}
  {[Online]. Available: http://arxiv.org/abs/quant-ph/9906054}\BibitemShut
  {NoStop}%
\bibitem [{\citenamefont {Bravyi}\ \emph
  {et~al.}(2018{\natexlab{a}})\citenamefont {Bravyi}, \citenamefont {Gosset},\
  and\ \citenamefont {K{\"{o}}nig}}]{Bravyi-science18}%
  \BibitemOpen
  \bibfield  {author} {\bibinfo {author} {\bibfnamefont {S.}~\bibnamefont
  {Bravyi}}, \bibinfo {author} {\bibfnamefont {D.}~\bibnamefont {Gosset}},\
  and\ \bibinfo {author} {\bibfnamefont {R.}~\bibnamefont {K{\"{o}}nig}},\
  }\bibfield  {title} {\bibinfo {title} {{Quantum advantage with shallow
  circuits.}},\ }\href {http://www.ncbi.nlm.nih.gov/pubmed/30337404} {\bibfield
   {journal} {\bibinfo  {journal} {Science}\ }\textbf {\bibinfo {volume}
  {362}},\ \bibinfo {pages} {308} (\bibinfo {year}
  {2018}{\natexlab{a}})}\BibitemShut {NoStop}%
\bibitem [{\citenamefont {Linke}\ \emph {et~al.}(2017)\citenamefont {Linke},
  \citenamefont {Maslov}, \citenamefont {Roetteler}, \citenamefont {Debnath},
  \citenamefont {Figgatt}, \citenamefont {Landsman}, \citenamefont {Wright},\
  and\ \citenamefont {Monroe}}]{Linke-nas17}%
  \BibitemOpen
  \bibfield  {author} {\bibinfo {author} {\bibfnamefont {N.~M.}\ \bibnamefont
  {Linke}}, \bibinfo {author} {\bibfnamefont {D.}~\bibnamefont {Maslov}},
  \bibinfo {author} {\bibfnamefont {M.}~\bibnamefont {Roetteler}}, \bibinfo
  {author} {\bibfnamefont {S.}~\bibnamefont {Debnath}}, \bibinfo {author}
  {\bibfnamefont {C.}~\bibnamefont {Figgatt}}, \bibinfo {author} {\bibfnamefont
  {K.~A.}\ \bibnamefont {Landsman}}, \bibinfo {author} {\bibfnamefont
  {K.}~\bibnamefont {Wright}},\ and\ \bibinfo {author} {\bibfnamefont
  {C.}~\bibnamefont {Monroe}},\ }\bibfield  {title} {\bibinfo {title}
  {Experimental comparison of two quantum computing architectures},\ }\href
  {https://www.pnas.org/content/114/13/3305/} {\bibfield  {journal} {\bibinfo
  {journal} {Proceedings of the National Academy of Sciences}\ }\textbf
  {\bibinfo {volume} {114}},\ \bibinfo {pages} {3305} (\bibinfo {year}
  {2017})}\BibitemShut {NoStop}%
\bibitem [{\citenamefont {Bengtsson}\ \emph {et~al.}(2014)\citenamefont
  {Bengtsson}, \citenamefont {Blanchfield}, \citenamefont {Campbell},\ and\
  \citenamefont {Howard}}]{Bengtsson-jphysa14}%
  \BibitemOpen
  \bibfield  {author} {\bibinfo {author} {\bibfnamefont {I.}~\bibnamefont
  {Bengtsson}}, \bibinfo {author} {\bibfnamefont {K.}~\bibnamefont
  {Blanchfield}}, \bibinfo {author} {\bibfnamefont {E.}~\bibnamefont
  {Campbell}},\ and\ \bibinfo {author} {\bibfnamefont {M.}~\bibnamefont
  {Howard}},\ }\bibfield  {title} {\bibinfo {title} {{Order 3 symmetry in the
  Clifford hierarchy}},\ }\href
  {https://iopscience.iop.org/article/10.1088/1751-8113/47/45/455302/meta}
  {\bibfield  {journal} {\bibinfo  {journal} {J. Phys. A Math. Theor.}\
  }\textbf {\bibinfo {volume} {47}},\ \bibinfo {pages} {455302} (\bibinfo
  {year} {2014})},\ \bibinfo {note} {[Online]. Available:
  http://arxiv.org/abs/1407.2713}\BibitemShut {NoStop}%
\bibitem [{\citenamefont {Cui}\ \emph {et~al.}(2017)\citenamefont {Cui},
  \citenamefont {Gottesman},\ and\ \citenamefont {Krishna}}]{Cui-physreva17}%
  \BibitemOpen
  \bibfield  {author} {\bibinfo {author} {\bibfnamefont {S.~X.}\ \bibnamefont
  {Cui}}, \bibinfo {author} {\bibfnamefont {D.}~\bibnamefont {Gottesman}},\
  and\ \bibinfo {author} {\bibfnamefont {A.}~\bibnamefont {Krishna}},\
  }\bibfield  {title} {\bibinfo {title} {{Diagonal gates in the Clifford
  hierarchy}},\ }\href
  {https://journals.aps.org/pra/pdf/10.1103/PhysRevA.95.012329} {\bibfield
  {journal} {\bibinfo  {journal} {Phys. Rev. A}\ }\textbf {\bibinfo {volume}
  {95}},\ \bibinfo {pages} {012329} (\bibinfo {year} {2017})},\ \bibinfo {note}
  {[Online]. Available: http://arxiv.org/abs/1608.06596}\BibitemShut {NoStop}%
\bibitem [{\citenamefont {Zhou}\ \emph {et~al.}(2000)\citenamefont {Zhou},
  \citenamefont {Leung},\ and\ \citenamefont {Chuang}}]{Zhou-pra00}%
  \BibitemOpen
  \bibfield  {author} {\bibinfo {author} {\bibfnamefont {X.}~\bibnamefont
  {Zhou}}, \bibinfo {author} {\bibfnamefont {D.~W.}\ \bibnamefont {Leung}},\
  and\ \bibinfo {author} {\bibfnamefont {I.~L.}\ \bibnamefont {Chuang}},\
  }\bibfield  {title} {\bibinfo {title} {{Methodology for quantum logic gate
  construction}},\ }\href {https://link.aps.org/doi/10.1103/PhysRevA.62.052316}
  {\bibfield  {journal} {\bibinfo  {journal} {Phys. Rev. A}\ }\textbf {\bibinfo
  {volume} {62}},\ \bibinfo {pages} {052316} (\bibinfo {year}
  {2000})}\BibitemShut {NoStop}%
\bibitem [{\citenamefont {Dehaene}\ and\ \citenamefont {{De
  Moor}}(2003)}]{Dehaene-physreva03}%
  \BibitemOpen
  \bibfield  {author} {\bibinfo {author} {\bibfnamefont {J.}~\bibnamefont
  {Dehaene}}\ and\ \bibinfo {author} {\bibfnamefont {B.}~\bibnamefont {{De
  Moor}}},\ }\bibfield  {title} {\bibinfo {title} {{Clifford group, stabilizer
  states, and linear and quadratic operations over GF(2)}},\ }\href
  {https://doi.org/10.1103/PhysRevA.68.042318} {\bibfield  {journal} {\bibinfo
  {journal} {Phys. Rev. A}\ }\textbf {\bibinfo {volume} {68}},\ \bibinfo
  {pages} {042318} (\bibinfo {year} {2003})}\BibitemShut {NoStop}%
\bibitem [{\citenamefont {Calderbank}\ and\ \citenamefont
  {Jafarpour}(2010)}]{Calderbank-seta10}%
  \BibitemOpen
  \bibfield  {author} {\bibinfo {author} {\bibfnamefont {R.}~\bibnamefont
  {Calderbank}}\ and\ \bibinfo {author} {\bibfnamefont {S.}~\bibnamefont
  {Jafarpour}},\ }\bibfield  {title} {\bibinfo {title} {{Reed Muller Sensing
  Matrices and the LASSO}},\ }in\ \href
  {https://link.springer.com/content/pdf/10.1007/978-3-642-15874-2.pdf#page=452}
  {\emph {\bibinfo {booktitle} {Intl. Conf. on Seq. Appl.}}}\ (\bibinfo
  {organization} {Springer},\ \bibinfo {year} {2010})\ pp.\ \bibinfo {pages}
  {442--463},\ \bibinfo {note} {[Online]. Available:
  http://arxiv.org/abs/1004.4949}\BibitemShut {NoStop}%
\bibitem [{\citenamefont {Tirkkonen}\ \emph {et~al.}(2017)\citenamefont
  {Tirkkonen}, \citenamefont {Boyd},\ and\ \citenamefont
  {Vehkalahti}}]{Tirkkonen-isit17}%
  \BibitemOpen
  \bibfield  {author} {\bibinfo {author} {\bibfnamefont {O.}~\bibnamefont
  {Tirkkonen}}, \bibinfo {author} {\bibfnamefont {C.}~\bibnamefont {Boyd}},\
  and\ \bibinfo {author} {\bibfnamefont {R.}~\bibnamefont {Vehkalahti}},\
  }\bibfield  {title} {\bibinfo {title} {{Grassmannian codes from multiple
  families of mutually unbiased bases}},\ }in\ \href
  {http://ieeexplore.ieee.org/document/8006636/} {\emph {\bibinfo {booktitle}
  {Proc.\ IEEE Int.\ Symp.\ Inform.\ Theory}}}\ (\bibinfo {year} {2017})\ pp.\
  \bibinfo {pages} {789--793}\BibitemShut {NoStop}%
\bibitem [{\citenamefont {Bravyi}\ and\ \citenamefont
  {Kitaev}(2005)}]{Bravyi-pra05}%
  \BibitemOpen
  \bibfield  {author} {\bibinfo {author} {\bibfnamefont {S.}~\bibnamefont
  {Bravyi}}\ and\ \bibinfo {author} {\bibfnamefont {A.}~\bibnamefont
  {Kitaev}},\ }\bibfield  {title} {\bibinfo {title} {{Universal quantum
  computation with ideal Clifford gates and noisy ancillas}},\ }\href
  {https://link.aps.org/doi/10.1103/PhysRevA.71.022316} {\bibfield  {journal}
  {\bibinfo  {journal} {Phys. Rev. A}\ }\textbf {\bibinfo {volume} {71}},\
  \bibinfo {pages} {022316} (\bibinfo {year} {2005})}\BibitemShut {NoStop}%
\bibitem [{\citenamefont {Gottesman}(2009)}]{Gottesman-arxiv09}%
  \BibitemOpen
  \bibfield  {author} {\bibinfo {author} {\bibfnamefont {D.}~\bibnamefont
  {Gottesman}},\ }\bibfield  {title} {\bibinfo {title} {{An Introduction to
  Quantum Error Correction and Fault-Tolerant Quantum Computation}},\
  }\href@noop {} {\bibfield  {journal} {\bibinfo  {journal} {arXiv preprint
  arXiv:0904.2557}\ } (\bibinfo {year} {2009})},\ \bibinfo {note} {[Online].
  Available: http://arxiv.org/abs/0904.2557}\BibitemShut {NoStop}%
\bibitem [{\citenamefont {Rengaswamy}\ \emph {et~al.}(2018)\citenamefont
  {Rengaswamy}, \citenamefont {Calderbank}, \citenamefont {Kadhe},\ and\
  \citenamefont {Pfister}}]{Rengaswamy-isit18}%
  \BibitemOpen
  \bibfield  {author} {\bibinfo {author} {\bibfnamefont {N.}~\bibnamefont
  {Rengaswamy}}, \bibinfo {author} {\bibfnamefont {R.}~\bibnamefont
  {Calderbank}}, \bibinfo {author} {\bibfnamefont {S.}~\bibnamefont {Kadhe}},\
  and\ \bibinfo {author} {\bibfnamefont {H.~D.}\ \bibnamefont {Pfister}},\
  }\bibfield  {title} {\bibinfo {title} {Synthesis of logical {C}lifford
  operators via symplectic geometry},\ }in\ \href
  {https://ieeexplore.ieee.org/document/8437652} {\emph {\bibinfo {booktitle}
  {Proc.\ IEEE Int.\ Symp.\ Inform.\ Theory}}}\ (\bibinfo {year} {2018})\ pp.\
  \bibinfo {pages} {791--795},\ \bibinfo {note} {[Online]. Available:
  http://arxiv.org/abs/1803.06987}\BibitemShut {NoStop}%
\bibitem [{\citenamefont {Gidney}\ and\ \citenamefont
  {Fowler}(2018)}]{Gidney-arxiv18}%
  \BibitemOpen
  \bibfield  {author} {\bibinfo {author} {\bibfnamefont {C.}~\bibnamefont
  {Gidney}}\ and\ \bibinfo {author} {\bibfnamefont {A.~G.}\ \bibnamefont
  {Fowler}},\ }\bibfield  {title} {\bibinfo {title} {{Efficient magic state
  factories with a catalyzed {\textbar}CCZ{\textgreater} to
  2{\textbar}T{\textgreater} transformation}},\ }\href@noop {} {\bibfield
  {journal} {\bibinfo  {journal} {arXiv preprint arXiv:1812.01238}\ } (\bibinfo
  {year} {2018})},\ \bibinfo {note} {[Online]. Available:
  http://arxiv.org/abs/1812.01238}\BibitemShut {NoStop}%
\bibitem [{\citenamefont {Can}(2018)}]{Can-sen18}%
  \BibitemOpen
  \bibfield  {author} {\bibinfo {author} {\bibfnamefont {T.}~\bibnamefont
  {Can}},\ }\emph {\bibinfo {title} {The {H}eisenberg-{W}eyl Group, Finite
  Symplectic Geometry, and their Applications}},\ \href@noop {} {\bibinfo
  {type} {{S}enior {T}hesis}},\ \bibinfo  {school} {Duke University} (\bibinfo
  {year} {2018})\BibitemShut {NoStop}%
\bibitem [{\citenamefont {Bravyi}\ \emph
  {et~al.}(2018{\natexlab{b}})\citenamefont {Bravyi}, \citenamefont {Browne},
  \citenamefont {Calpin}, \citenamefont {Campbell}, \citenamefont {Gosset},\
  and\ \citenamefont {Howard}}]{Bravyi-arxiv18}%
  \BibitemOpen
  \bibfield  {author} {\bibinfo {author} {\bibfnamefont {S.}~\bibnamefont
  {Bravyi}}, \bibinfo {author} {\bibfnamefont {D.}~\bibnamefont {Browne}},
  \bibinfo {author} {\bibfnamefont {P.}~\bibnamefont {Calpin}}, \bibinfo
  {author} {\bibfnamefont {E.}~\bibnamefont {Campbell}}, \bibinfo {author}
  {\bibfnamefont {D.}~\bibnamefont {Gosset}},\ and\ \bibinfo {author}
  {\bibfnamefont {M.}~\bibnamefont {Howard}},\ }\bibfield  {title} {\bibinfo
  {title} {{Simulation of quantum circuits by low-rank stabilizer
  decompositions}},\ }\href@noop {} {\bibfield  {journal} {\bibinfo  {journal}
  {arXiv preprint arXiv:1808.00128}\ } (\bibinfo {year}
  {2018}{\natexlab{b}})},\ \bibinfo {note} {[Online]. Available:
  http://arxiv.org/abs/1808.00128}\BibitemShut {NoStop}%
\bibitem [{\citenamefont {Calderbank}\ \emph {et~al.}(1998)\citenamefont
  {Calderbank}, \citenamefont {Rains}, \citenamefont {Shor},\ and\
  \citenamefont {Sloane}}]{Calderbank-it98*2}%
  \BibitemOpen
  \bibfield  {author} {\bibinfo {author} {\bibfnamefont {R.}~\bibnamefont
  {Calderbank}}, \bibinfo {author} {\bibfnamefont {E.}~\bibnamefont {Rains}},
  \bibinfo {author} {\bibfnamefont {P.}~\bibnamefont {Shor}},\ and\ \bibinfo
  {author} {\bibfnamefont {N.}~\bibnamefont {Sloane}},\ }\bibfield  {title}
  {\bibinfo {title} {Quantum error correction via codes over {GF}(4)},\ }\href
  {https://ieeexplore.ieee.org/document/613213} {\bibfield  {journal} {\bibinfo
   {journal} {IEEE Trans.\ Inform.\ Theory}\ }\textbf {\bibinfo {volume}
  {44}},\ \bibinfo {pages} {1369} (\bibinfo {year} {1998})}\BibitemShut
  {NoStop}%
\bibitem [{\citenamefont {Gottesman}(1997)}]{Gottesman-phd97}%
  \BibitemOpen
  \bibfield  {author} {\bibinfo {author} {\bibfnamefont {D.}~\bibnamefont
  {Gottesman}},\ }\emph {\bibinfo {title} {Stabilizer codes and quantum error
  correction}},\ \href@noop {} {Ph.D. thesis},\ \bibinfo  {school} {California
  Institute of Technology} (\bibinfo {year} {1997})\BibitemShut {NoStop}%
\bibitem [{\citenamefont {Wilde}(2009)}]{Wilde-physreva09}%
  \BibitemOpen
  \bibfield  {author} {\bibinfo {author} {\bibfnamefont {M.~M.}\ \bibnamefont
  {Wilde}},\ }\bibfield  {title} {\bibinfo {title} {{Logical operators of
  quantum codes}},\ }\href {https://doi.org/10.1103/PhysRevA.79.062322}
  {\bibfield  {journal} {\bibinfo  {journal} {Phys. Rev. A}\ }\textbf {\bibinfo
  {volume} {79}},\ \bibinfo {pages} {062322} (\bibinfo {year}
  {2009})}\BibitemShut {NoStop}%
\end{thebibliography}


%

\end{document}